\documentclass[showpacs,preprintnumbers,amsmath,amssymb]{revtex4}

\usepackage{graphicx}
\usepackage{dcolumn}
\usepackage{bm}
\usepackage{epsf}

\newcommand{\eq}[1]{eq.~(\ref{#1})}
\newcommand{\eqs}[2]{eqs.(\ref{#1},\ref{#2})}

\newcommand{\eqsss}[2]{eqs.(\ref{#1}--\ref{#2})}
\newcommand{\Eq}[1]{Eq.~(\ref{#1})}
\newcommand{\Eqs}[2]{Eqs.(\ref{#1},\ref{#2})}

\newcommand{\ur}[1]{(\ref{#1})}
\newcommand{\urs}[2]{(\ref{#1},\ref{#2})}

\newcommand{\beq}{\begin{equation}}
\newcommand{\eeq}{\end{equation}}

\newcommand{\la}[1]{\label{#1}}
\newcommand{\bea}{\begin{eqnarray}}
\newcommand{\eea}{\end{eqnarray}}
\newcommand{\ba}{\begin{array}}
\newcommand{\ea}{\end{array}}

\newcommand{\half}{{\textstyle{\frac{1}{2}}}}
\newcommand{\noi}{\noindent}

\newcommand{\n}{\nonumber}

\newcommand{\Tr}{{\rm Tr}}

\begin{document}

\title{Confining ensemble of dyons}
\author{Dmitri Diakonov}
\email{diakonov@thd.pnpi.spb.ru}
\author{Victor Petrov}
\email{victorp@thd.pnpi.spb.ru}
\vskip 0.3true cm

\affiliation{St. Petersburg NPI, Gatchina, 188 300, St. Petersburg, Russia}

\date{June 20, 2007}

\begin{abstract}
We construct the integration measure over the moduli space of an arbitrary
number of $N$ kinds of dyons of the pure $SU(N)$ gauge theory at finite
temperatures. The ensemble of dyons governed by the measure is mathematically
described by a (supersymmetric) quantum field theory that is exactly solvable
and is remarkable for a number of striking features: 1) The free energy has the
minimum corresponding to the zero average Polyakov line, as expected in the
confining phase; 2)~The correlation function of two Polyakov lines exhibits a
linear potential between static quarks in any $N$-ality non-zero
representation, with a calculable string tension roughly independent of
temperature; 3) The average spatial Wilson loop falls off exponentially with
its area and the same string tension; 4) At a critical temperature the ensemble
of dyons rearranges and de-confines; 5)~The estimated ratio of the critical
temperature to the square root of the string tension is in excellent agreement
with the lattice data.
\end{abstract}




\maketitle

\tableofcontents

\section{Introduction}

Isolated dyons in the pure Yang--Mills theory are (anti) self-dual solutions
of the equation of motion $D_\mu F_{\mu\nu}=0$, which in an appropriate gauge
are Abelian at large distances from the centers and carry unity electric and
magnetic charges with respect to the Cartan generators $C_m$:
\beq
\pm{\bf E}_m={\bf B}_m=\frac{1}{2}\frac{{\bf r}}{|{\bf r}|^3}\,C_m.
\la{EB}\eeq
For the $SU(N)$ gauge group on which we focus in this paper
there are $N-1$ Cartan generators associated with the simple roots of the
group, $C_m={\rm diag}(0,...,1,-1,0,...,0)$ where the $1$ is on the $m^{\rm th}$
place, that are supplemented by the $N^{\rm th}$ generator $C_N={\rm diag}(-1,0,...,1)$
to make the set of $N$ dyons electric- and magnetic-neutral. The first $N\!-\!1$ dyons
are also called the Bogomolny--Prasad--Sommerfield (BPS) monopoles~\cite{BPS}.
The last $N^{\rm th}$ dyon is sometimes called the Kaluza--Klein monopole:
in the gauge where the first $N\!-\!1$ dyons are described by a static field
the last has time-dependent fields inside the core. However, by a periodic
time-dependent gauge transformation one can make the last one time-independent
(at the cost of a time dependence inside other dyons), therefore the distinction
of the last dyon is illusory and we shall treat all $N$ of them on the same footing.
The cyclic symmetry of $N$ dyons is evident from the $D$-branes point of
view~\cite{LeeYi,DHKM-99}.

In this paper we explore the properties of a semiclassical vacuum built of a
large number of dyons of $N$ kinds.

To make the semiclassical calculation of the Yang--Mills partition function
well defined one needs {\it i}) to expand about a true saddle point of the
action, {\it ii}) to be sure that the quantum fluctuation determinant is
infrared-finite. For an isolated dyon the first is true but the second is
false. For an arbitrary superposition of $N$ different-kind dyons the second
is true but the first is wrong. To satisfy both requirements one can consider
$N$ dyons as constituents of the Kraan--van Baal--Lee--Lu (KvBLL) instantons
with non-trivial holonomy~\cite{KvB,LL}, which are saddle points of the Yang--Mills
partition function as they are exact self-dual solutions of the equations of
motion. At the same time the fluctuation determinant about the KvBLL instanton
is infrared-finite (and actually exactly calculable~\cite{DGPS}) since its total
electric and magnetic charges are zero.

KvBLL instantons generalize the standard Belavin--Polyakov--Schwartz--Tuypkin (BPST)
instantons~\cite{BPST} having trivial holonomy. The mere notion of dyons and of the KvBLL
instantons (also called calorons) alike imply that the Yang--Mills field is periodic in
the Euclidean time direction, as it is in the case of non-zero temperature. Therefore
we shall be considering the Yang--Mills partition function at finite temperature.
However, the circumference of the compactified space can be gradually put to infinity,
corresponding to the zero temperature limit. In that limit, the temperature can be
considered as an infrared regulator of the theory needed to distinguish between
the trivial and non-trivial holonomy. After all the temperature of the Universe is 2.7 K $\neq 0$.

In this context, the holonomy is defined as the set $\{\mu_m\}$ of the gauge-invariant
eigenvalues of the Polyakov loop $L$ winding in the compactified time direction,
at spatial infinity:
\beq
L={\rm P}\,\exp\left(i\int_0^{1/T}\!\!dt\,A_4\right)_{|\vec  x|\to\infty}
=V\,{\rm diag}\left(e^{2\pi i \mu_1},\,e^{2\pi i \mu_2},\ldots,e^{2\pi i \mu_N}\right)\,V^{-1},
\qquad \sum_{m=1}^N\mu_m=0.
\la{Pol1}\eeq
By making a global gauge rotation one can always order the eigenvalues such that
\beq
\mu_1\leq \mu_2\leq \ldots \leq \mu_N\leq \mu_{N+1}\equiv \mu_1+1.
\la{mus}\eeq
which we shall assume done. If all eigenvalues are equal up to an integer,
implying
\beq
\mu_m^{\rm triv}=\left\{\begin{array}{c} \frac{k}{N}-1\;\;  {\rm when}\; m\leq k,\\
\frac{k}{N}\;\; {\rm when}\; m > k \end{array}\right.,\quad {\rm where}\; k=1,...,N,
\la{mutriv}\eeq
the Polyakov line belongs to the $SU(N)$ group center, and the holonomy is then said
to be ``trivial'': $L_{\rm triv}={\rm diag}\left(\exp(2\pi i k/N),...,\exp(2\pi i k/N)\right),
\;k=1,...,N$. Standard BPST instantons, as well as their genuine periodic generalization
to non-zero temperatures by Harrington and Shepard~\cite{HS}, possess trivial holonomy,
whereas for the KvBLL instantons the gauge-invariant eigenvalues of the Polyakov line
assume, generally, non-equal values corresponding to a ``non-trivial'' holonomy.
Among these, there is a special set of equidistant $\mu_m$'s that can be named a
``maximally non-trivial'' holonomy,
\beq
\mu_m^{\rm conf}=-\frac{1}{2}-\frac{1}{2N}+\frac{m}{N},
\la{muconf}\eeq
having a distinguished property that it leads to $\Tr\,L=0$. Since the average
Polyakov line is zero in the confining phase, the set \ur{muconf} can be also
called the ``confining'' holonomy.

Whatever the set of $\mu$'s is equal to, it is a global characterization of the
Yang--Mills system. Integrating over all possible $\mu$'s is equivalent to
requesting that the total color charge of the system is zero~\cite{GPY}. Since
$\mu$'s are constants, the ultimate partition function has to be extensive in
these quantities, meaning ${\cal Z}=\exp[-F(\{\mu\},T)V]$ where $F$ is
the free energy and $V$ is the 3-volume. If $F(\{\mu\})$ has a minimum
for a particular set $\{\mu\}$, integration over $\mu$'s is done by saddle
point method justified in the thermodynamic limit $V\to\infty$, hence the Yang--Mills
system settles at the minimum of the free energy as function of $\mu$'s. The
big question is whether the pure Yang--Mills theory has the minimum of the
free energy at the ``confining'' holonomy \ur{muconf}, or elsewhere.

It has been argued long ago~\cite{GPY} that the pure Yang--Mills theory has $N$
minima of the free energy at the trivial holonomy \ur{mutriv}. To that end, one
refers to the perturbative potential energy as function of spatially constant
$A_4$~\cite{GPY,NW}:
\beq
P^{\rm pert}=V\,\frac{(2\pi)^2T^3}{3}\sum_{m>n}^N(\mu_m-\mu_n)^2[1-(\mu_m-\mu_n)]^2,
\la{Ppert}\eeq
which, indeed, has $N$ minima (all with zero energy) at the trivial holonomy
\ur{mutriv}, corresponding to $N$ elements of the center of $SU(N)$. The
``confining'' holonomy \ur{muconf} corresponds, on the contrary, to the non-degenerate
maximum of \ur{Ppert}, equal to
\beq
P^{\rm pert,\,max}=V\,\frac{(2\pi)^2T^3}{180}\frac{N^4-1}{N^2}.
\la{Ppertmax}\eeq
The large volume factor in \eq{Ppert} seemingly prohibits any configurations with non-trivial
holonomy, dyons and KvBLL instantons included.

A loophole in this dyon-killing argument has been noticed in Ref.~\cite{D02}:
If one takes an {\em ensemble} of dyons, with their number proportional to the volume,
the non-perturbative dyon-induced potential energy is also proportional to the volume
and may hence override the perturbative one, possibly leading to another minimum
of the full free energy. This scenario was made probable in Ref.~\cite{DGPS}
where it was shown that the non-perturbative potential energy induced by a dilute gas
of the KvBLL instantons prevailed over the perturbative one at temperatures below some
critical $T_c$ estimated through $\Lambda$, the Yang--Mills scale parameter, and that
the trivial holonomy was not the minimum of the full free energy anymore. Below that
critical temperature the KvBLL instantons dissociate into individual dyons. The
problem, therefore, is to build the partition function for dissociated dyons, and to
check if the full free energy has a minimum at the confining holonomy \ur{muconf}.
We get an affirmative answer to this question.

The moduli space of a single KvBLL instanton is characterized by $4N$ parameters
(of which the classical action is independent); these can be conveniently chosen as
$3d$ coordinates of $N$ dyons constituting the instanton, and their $U(1)$ phases,
$3N+N=4N$. In the part of the moduli space where all dyons are well separated,
the KvBLL instanton becomes a sum of $N$ types of BPS monopoles with a time-independent
action density. At small separations between dyons the action density of the KvBLL
instanton is time-dependent and resembles that of the standard BPST instanton.
The KvBLL instanton reduces to the standard BPST instanton in the two limiting cases:
i) trivial holonomy (all $\mu$'s are equal up to an integer) and any temperature,
ii) non-trivial holonomy but vanishing temperature, provided the separations between
dyons shrink to zero as $\sim \rho^2T$ where $\rho$ is the standard instanton size.

The quantum weight of the KvBLL instanton is determined by a product of two
factors: {\it i}) the determinant of the moduli space metric, {\it ii}) the
small-oscillation determinant over non-zero modes in the KvBLL background.
The latter has been computed exactly in Ref.~\cite{DGPS} for the $SU(2)$ gauge group;
recently the result has been generalized to any $SU(N)$~\cite{GS-SUN}.
The former is also known exactly (see the references and discussion in the next
Section). These achievements, however, are limited to the case of a single KvBLL
instanton with unity topological charge. To build the dyon vacuum, one needs
multi-instanton solutions, with the topological charge proportional to the
volume, similar to the case of the instanton liquid model~\cite{ILM,D02}.
Although there has been serious progress recently in constructing general
multi-instanton solutions with non-trivial holonomy and their moduli space
metric~\cite{Nogradi}, a desirable explicit construction is still lacking.

Nevertheless, the moduli space metric can be constructed by combining the
metric known for $N$ different-kind dyons of the $SU(N)$ group with another
known metric for same-kind dyons, and by taking into due account the permutational
symmetry between identical dyons. One of the two ingredients, the metric for
different-kind dyons, is known exactly for all separations and involves only
Coulomb-like interactions. The other ingredient related to the same-kind dyons
is more complex. The metric for any separations between same-kind dyons
allows for charge exchange between dyons and involves elliptic functions of separation.
For two dyons of same charge the exact metric was found by Atiyah and Hitchin~\cite{AH}
from the requirement that the Riemann tensor constructed from the metric must
be self-dual. When the separation between same-kind dyons exceeds their core
sizes, charge exchange dies out exponentially with the separation, the metric becomes
simple and can be written for any number of same-kind dyons with the exponential
precision~\cite{GM1,GM2}. It involves only the Coulomb-like interactions and is in fact
very similar to that for different-kind dyons but with opposite signs in the
Coulomb bonds. It is this marriage of the asymptotic form of the metric for
same-kind dyons, valid with exponential precision, with the metric for different-kind
dyons, valid for {\em any} separations, that we shall explore in this paper.

In fact, it may prove to be sufficient for an accurate description of the ensemble
of dyons in the thermodynamic limit ($V\to\infty$), as dyons of the same kind repulse
each other whereas dyons of different kind attract each other. Therefore small separations
between same-kind dyons, where our measure is only approximate, may be statistically
unimportant. [Unfortunately, taking the large $N$ limit does not help since only
nearest neighbors in color interact, hence at any $N>2$ there are only twice more
different-kind bonds than same kind.] Indeed, we find that, despite an approximate
integration measure, the Lorentz symmetry is, in a sense, restored at $T\to 0$:
the ``electric'' string tension as determined from the correlation of Polyakov lines
turns out to be independent of temperature and {\em equal} to the ``magnetic'' string
tension determined from the area law for spatial Wilson loops, for all
representations considered. The free energy itself also has a reasonable limit at
$T\to 0$. However, to remain on the safe side, we claim the results
only for sufficiently high temperatures (but below the deconfinement phase transition)
where dyons do not overlap on the average, and the metric used is justified.

This study is exploratory as we ignore many essential ingredients of the full
Yang--Mills theory. In particular, we consider only the ensemble of dyons of
one duality, whereas $CP$ invariance of the vacuum requires that there must be
an equal number of self-dual and anti-self-dual configurations, up to the
thermodynamic fluctuations $\sim \sqrt{V}$~\cite{ILM}. We basically ignore the
determinant over non-zero modes, taking from it only certain known salient
features, like the renormalization of the coupling constant and the
perturbative potential energy \ur{Ppert}. Our aim is to demonstrate that the
integration measure over dyons has a drastic, probably a decisive effect on the
ensemble of dyons, that the ensemble can be mathematically described by an
exactly solvable field theory in three dimensions, and that the resulting
vacuum built of dyons has certain features expected for the confining pure
Yang--Mills theory.

A more phenomenological and lattice-oriented attempt to describe the
ensemble of the KvBLL instantons, has been proposed recently in Ref.~\cite{Ilgenfritz}.

\section{Integration measure over dyons}

\subsection{Different-kind dyons}

The metric of the moduli space of a single $SU(N)$ KvBLL instanton, written in terms
of $N$ different-kind dyons' coordinates and $U(1)$ phases, was first conjectured by Lee,
Weinberg and Yi~\cite{LWY} generalizing the previous work by Gibbons and Manton~\cite{GM2},
and then confirmed by Kraan~\cite{Kraan} by an explicit calculation of the zero-mode Jacobian
from the Atiyah--Drinfeld--Hitchin--Manin--Nahm (ADHMN) construction~\cite{ADHM,Nahm80}
for the $SU(N)$ caloron~\cite{KvBSUN}. It was later checked independently in Ref.~\cite{DG05},
also by an explicit calculation of the Jacobian.

There are several equivalent ways to present the metric of a single $SU(N)$ KvBLL
instanton; we use here the form suggested originally by Gibbons and Manton~\cite{GM2}
(although these authors considered another case -- that of same kind dyons, see the
next subsection):
\beq
ds^2=G_{mn}\,d{\bf x}_m\cdot d{\bf x}_n+(d\psi_m+{\bf W}_{mm'}\cdot d{\bf x}_{m'})
\,G^{-1}_{mn}\,(d\psi_n+{\bf W}_{nn'}\cdot d{\bf x}_{n'}),\qquad m,n=1...N.
\la{m1}\eeq
Here ${\bf x}_m,\,m=1...N,$ are the $3d$ centers of dyons, and $\psi_m$ are
their $U(1)$ phases. $G_{mn}$ is a symmetric matrix composed of the Coulomb
interactions between dyons that are nearest ``neighbors in color'':
\beq
G_{mn}=\delta_{mn}\,\left(4\pi\nu_m+\frac{1}{|{\bf x}_m-{\bf x}_{m-1}|}
+\frac{1}{|{\bf x}_m-{\bf x}_{m+1}|}\right)-\frac{\delta_{m,n-1}}{|{\bf x}_m-{\bf x}_{m+1}|}
-\frac{\delta_{m,n+1}}{|{\bf x}_m-{\bf x}_{m-1}|},
\la{G1}\eeq
where $\nu_m=\mu_{m+1}-\mu_m;\;\nu_1+...+\nu_N=1$ (see Section I). Periodicity in color indices
is implied throughout the paper: $m=N+1$ is equivalent to $m=1$, $m=0$ is equivalent to $m=N$.
${\bf W}_{mn}$ are three $N\times N$ symmetric matrices composed of the
electric charge--magnetic charge interaction potential ${\bf w}({\bf x})$:
\beq
{\bf W}_{mn}=\delta_{mn}\,\left({\bf w}({\bf x}_m-{\bf x}_{m-1})
+{\bf w}({\bf x}_m-{\bf x}_{m+1})\right)
-\delta_{m,n-1}{\bf w}({\bf x}_m-{\bf x}_{m+1})
-\delta_{m,n+1}{\bf w}({\bf x}_m-{\bf x}_{m-1})
\la{W1}\eeq
where ${\bf w}({\bf x})$ satisfies the equation $\epsilon^{abc}\partial_b{\bf w}_c
=-{\bf x}_a/|{\bf x}|^3$; to solve it, one has to introduce a Dirac string
singularity. Choosing the string, {\it e.g.,} along the third axis and parameterizing
the separation vector between two dyons in spherical coordinates ${\bf x}
=|{\bf x}|(\sin\theta\,\cos\phi,\,\sin\theta\,\sin\phi,\,\cos\theta)$
one finds ${\bf w}({\bf x})=(-\cot\theta\,\sin\phi,\,\cot\theta\,\cos\phi,\,0)/|{\bf x}|$,
such that the factor $(d\psi+{\bf w}\cdot d{\bf x})$ written for relative
coordinates combines into
\beq
d\Sigma_3=d\psi+\cos\theta\,d\phi
\la{Sigma3}\eeq
which is a familiar 1-form encountered, {\it e.g.}, in the theory of rigid-body
rotations, being the third projection of the angular velocity.
The fact that this quantity arises in the context of electric charge--magnetic charge
interaction, was known for quite a while~\cite{Guadagnini}. A more general,
basis-independent way to present the interaction is via the Wess--Zumino
term~\cite{DPmon}.

The temperature factors $T$ have been dropped in eqs.({\ref{m1}}-{\ref{W1}})
but can be anytime restored from dimensions. We stress that the metric \ur{m1} is exact
and valid for {\em any} $3d$ separations between dyons inside the KvBLL
instanton, including the case when they strongly overlap. For more details on
this metric see Refs.~\cite{Kraan,DG05}.

The integration measure for the KvBLL instanton is
\beq
\prod_{m=1}^N d^3{\bf x}_m d\psi_m\,\sqrt{\det g},\qquad  \sqrt{\det g}=\det G,
\la{me1}\eeq
where $g$ is the full $4N\times 4N$ metric tensor given by \eq{m1}. In computing $\det g$
one can shift $d\psi_m\to d\psi'_m=d\psi_m +{\bf W}_{mk}\cdot d{\bf x}_k$, therefore,
$\det g=(\det G)^3\,\det G^{-1}= (\det G)^2$, hence $\sqrt{\det g}=\det G$ where
$G$ is given by \eq{G1}~\cite{DG05}. In that reference it was also shown that
in the limit of trivial holonomy or small temperatures, the integration measure \ur{me1}
reduces to the well-known one for the standard BPST instanton~\cite{Bernard}.

\subsection{Same-kind dyons}

For multi-KvBLL instantons, a new element appears, namely two or more {\em same-kind}
dyons are present. For two dyons of the same kind, the metric splits into a flat
$4d$ space for center-of-mass coordinates and a non-flat $4d$ space $M_2$ for relative
coordinates ${\bf r}$ and $\psi$. Self-duality implies that $M_2$ is a self-dual
Einstein manifold. Gibbons and Pope~\cite{GP} proposed the following form for the metric:
\beq
ds^2=f^2\,dr^2+a^2\,d\Sigma_1^2+b^2\,d\Sigma_2^2+c^2\,d\Sigma_3^2,\qquad
\sqrt{g}=|f a b c|\sin\theta,
\la{m2}\eeq
where
\bea\n
d\Sigma_1&=&-\sin\psi\,d\theta+\cos\psi\,\sin\theta\,d\phi,\\
\n
d\Sigma_2&=&\cos\psi\,d\theta+\sin\psi\,\sin\theta\,d\phi,
\eea
and $d\Sigma_3$ is the third `angular velocity' \ur{Sigma3}; a,b,c,f
are functions of the dyon separation $r$. Self-duality requires that the Riemann
tensor built from the metric \ur{m2} satisfies
$R_{\alpha\beta\gamma\delta}=\half\,\sqrt{g}\,\epsilon_{\gamma\delta\kappa\lambda}\,
R_{\alpha\beta}^{\;\;\;\;\;\kappa\lambda}$ which leads to the system of first-order
equations
\beq
\frac{1}{f}\frac{da}{dr}=\frac{b^2+c^2-a^2}{2bc}-\lambda,\quad
{\rm and\;cyclic\;permutations\;of}\;a,b,c,
\la{abc}\eeq
where $\lambda=0$ or $1$. The value $\lambda=1$ is chosen from symmetry
considerations~\cite{AHbook}. Eqs.({\ref{abc}}) have a simple solution~\cite{GM1}
\beq
f=-\beta\sqrt{1+\frac{2\alpha}{r}},\quad
a=b=\beta\,r\,\sqrt{1+\frac{2\alpha}{r}},\quad
c=\frac{2\alpha\beta}{\sqrt{1+\frac{2\alpha}{r}}}
\la{sol0}\eeq
with any $\alpha,\beta$. We fix $\alpha,\beta$ from the asymptotics of
the metric of two dyons of the same kind $m$ at $r\to\infty$:
$\beta=\sqrt{2\pi\nu_m},\;\alpha=-1/(2\pi\nu_m)$~\cite{footnote1}. To get the
full $8\times 8$ metric tensor in terms of the two dyons' coordinates
\bea\n
&&{\bf x}_1={\bf X}+\half {\bf r},\quad d\psi_1=d\Psi+\half d\psi,\\
\n
&&{\bf x}_2={\bf X}-\half {\bf r},\quad d\psi_2=d\Psi-\half d\psi,
\eea
we add to \eq{m2} the flat metric for center-of-mass coordinates
$8\pi\nu(d{\bf X}\!\cdot\!d{\bf X}+d\Psi\,d\Psi)$ and obtain the metric which
is very similar to \eq{m1}:
\bea\la{m21}
ds^2&=&\tilde G_{ij}\,d{\bf x}_i\!\cdot\!d{\bf x}_j
+(d\psi_i+\tilde{\bf W}_{ii'}\!\cdot\!d{\bf x}_{i'})\,\tilde G^{-1}_{ij}\,
(d\psi_j+\tilde{\bf W}_{jj'}\!\cdot\!d{\bf x}_{j'}),
\quad i,j=1,2,\\
\la{G21}
\tilde G_{ij}&=& \left(\begin{array}{cc}4\pi\nu_m-\frac{2}{|{\bf x}_1-{\bf x}_2|}
& \frac{2}{|{\bf x}_1-{\bf x}_2|}\\
\frac{2}{|{\bf x}_1-{\bf x}_2|} & 4\pi\nu_m-\frac{2}{|{\bf x}_1-{\bf x}_2|}
\end{array}\right),\\
\la{W21}\tilde W_{ij} &=& \left(\begin{array}{cc}-{\bf w}({\bf x}_1\!-\!{\bf x}_2) &
{\bf w}({\bf x}_1\!-\!{\bf x}_2)\\ {\bf w}({\bf x}_1\!-\!{\bf x}_2) &
-{\bf w}({\bf x}_1\!-\!{\bf x}_2)\end{array}\right),\qquad
{\bf w}({\bf x})=\frac{1}{|{\bf x}|}\,(-\cot\theta\,\sin\phi,\,\cot\theta\,\cos\phi,\,0).
\eea
Note the opposite sign of Coulomb interactions in \eq{G21} as compared to
\eq{G1}.

It is easy to generalize eqs.(\ref{m21}-\ref{W21}) to {\em any} number of
dyons, all of the $m^{\rm th}$ kind. One extends the summation in \eq{m21} from
$i,j\leq 2$ to $i,j\leq K$, where $K$ is the number of same-kind dyons, and modifies \eqs{G21}{W21} as
\beq\la{GW22}
\tilde G_{ij}=\left\{\begin{array}{cc}4\pi\nu_m-\sum_{k\neq i}\frac{2}{|{\bf x}_i-{\bf
x}_k|},& i=j,\\ \frac{2}{|{\bf x}_i-{\bf x}_j|}, & i\neq
j\end{array}\right.,\qquad
\tilde W_{ij}=\left\{\begin{array}{cc}-2\sum_{k\neq i}{\bf w}({\bf x}_i
-{\bf x}_k), & i=j,\\ 2\,{\bf w}({\bf x}_i-{\bf x}_j), & i\neq j\end{array}\right.\,.
\eeq
\Eqs{m21}{GW22} were derived by Gibbons and Manton~\cite{GM2} (with other coefficients
related to another scale convention) from considering the classical equations of motion
for $K$ identical monopoles at large separations.

Although \eq{sol0} from where the metric \ur{m21} stems, is an exact solution
of the Einstein self-duality equation, it is believed that \eq{m21} is valid
only for large separations $r>1/2\pi\nu_m T$ (we restore the explicit
temperature factor here). Note that a very similar metric \ur{m1} for
different-kind dyons is proven to be valid at any separations. A somewhat
superficial reason for the difference between same- and different-kind dyons
was noted in Ref.~\cite{LWY}: while the metric \ur{m1} is positive-definite,
the metric \ur{m21} goes to zero at $r=1/2\pi\nu_m T$.
A deeper reason is that while the metric \ur{m1} describes a system with total
electric and magnetic charges zero (the KvBLL instanton), the metric \ur{m21}
is applied to a system where neither is zero.

A non-trivial solution of the Einstein self-duality \eq{abc} was found by
Atiyah and Hitchin (AH)~\cite{AH} (for more details see~\cite{AHbook}). In the
AH solution, the $a,b,c$ functions are given by elliptic integrals, and
$a(r)\neq b(r)$. It follows then from \eq{m2} that the relative $U(1)$
orientation angle $\psi$ enters the metric in an essential way, in particular,
a shift of $\psi$ is not an isometry anymore. The functions $a,b,c$ of the
AH solution differ from those of the solution \ur{sol0} by terms of the
order of $\sim\!\exp(-4\pi\nu_m T\,r)$ which die out exponentially at large
separations~\cite{GM1}. Therefore, at large $r$ the AH solution takes the form
of \eq{sol0} such that $\psi$ enters the metric in a trivial way, as in \eq{m21}.

For the AH solution, the metric determinant goes to zero at even larger
$r=1/(4\nu_m T)$~\cite{AH,GM1}.
Physically, this point corresponds to an axially-symmetric two-monopole solution where
two monopoles coincide. When dyons overlap, what should be called ``separation''
becomes ambiguous; $r$ is defined only in the context of a concrete parameterization
of the field. In the ensemble, the zero of the metric determinant means a vanishing
contribution to the partition function, actually imposing a very strong repulsion
between same-kind dyons. The same is true for the approximate metric \ur{m21}
which we shall use below. One can think that the approximate metric \ur{m21},
because of the strong repulsion it imposes, strongly suppresses in the statistical
mechanics' sense configurations with close same-kind dyons, where the metric
becomes inaccurate. In other words, the approximate metric \ur{m21} may be accurate
for statistically important configurations. This hypothesis needs a detailed study,
of course. Its consequences, however, turn out to be reasonable.

\subsection{Combining the metric for same-kind and different-kind dyons}

The explicit form of the metric tensor for $K$ KvBLL instantons made of $N$ kinds
of dyons is not known (for the latest development see Ref.~\cite{Nogradi}).
Below we suggest an Ansatz for this metric, satisfying the known requirements.
One has to combine the metric \ur{m1} for $N$ different-kind dyons with that
for $K$ same-kind ones. The solution of the problem is almost obvious if one
takes the approximate metric \ur{m21} for same-kind dyons, as it has exactly
the same form as the metric \ur{m1} for different-kind dyons. Since the metric
cannot `know' to which instanton a particular dyon belongs to, it must be
symmetric under permutations of any pair of dyons of the same kind. Important,
the metric of the moduli space of self-dual solutions must be
hyper-K\"ahler~\cite{AHbook}.

Let indices $m,n=1...N$ refer to the dyon kind (or `color') and indices $i,j,k=1...K$
number the dyons of the same kind. The coordinates of the $i^{\rm th}$ dyon of the
$m^{\rm th}$ kind are $({\bf x}_{mi},\,\psi_{mi})\equiv y_{mi}^\alpha\equiv y_A^\alpha,
\;\alpha=1,2,3,4$. To shorten notations, we introduce instead of the multi-index $(mi)$
a single index $A=(mi)$ running from 1 to $KN$.

We write the full metric tensor as
\beq\la{m3}
ds^2=g_{A\alpha,B\beta}\,dy_A^\alpha dy_B^\beta=G_{AB}\,d{\bf
x}_A\!\cdot\!d{\bf x}_B +(d\psi_A+{\bf W}_{AA'}\!\cdot\!d{\bf x}_{A'})\,
G^{-1}_{AB}\,(d\psi_B+{\bf W}_{B,B'}\!\cdot\!d{\bf x}_{B'}),
\eeq
where, explicitly,
\bea\la{G3}
G_{AB}&=&G_{mi,nj}=\delta_{mn}\delta_{ij}\,\left(4\pi\nu_m+\sum_k\frac{1}{|{\bf
x}_{mi}\!-\!{\bf x}_{m-1,k}|} +\sum_k\frac{1}{|{\bf x}_{mi}\!-\!{\bf
x}_{m+1,k}|}
-2\sum_{k\neq i}\frac{1}{|{\bf x}_{mi}\!-\!{\bf x}_{mk}|}\right)\\
\n
&&-\frac{\delta_{m,n-1}}{|{\bf x}_{mi}\!-\!{\bf x}_{m+1,j}|}
-\frac{\delta_{m,n+1}}{|{\bf x}_{mi}\!-\!{\bf x}_{m-1,j}|}
+2\left.\frac{\delta_{mn}}{|{\bf x}_{mi}\!-\!{\bf x}_{mj}|}\right|_{i\neq j},
\eea
\bea\la{W3}
{\bf W}_{AB}&=&{\bf W}_{mi,nj}=\delta_{mn}\delta_{ij}\,\left(\sum_k{\bf w}({\bf x}_{mi}\!-\!{\bf x}_{m-1,k})
+\sum_k{\bf w}({\bf x}_{mi}\!-\!{\bf x}_{m+1,k})
-2\sum_{k\neq i}{\bf w}({\bf x}_{mi}\!-\!{\bf x}_{mk})\right)\\
\n
&&-\delta_{m,n-1}{\bf w}({\bf x}_{mi}\!-\!{\bf x}_{m+1,j})
-\delta_{m,n+1}{\bf w}({\bf x}_{mi}\!-\!{\bf x}_{m-1,j})
+2\delta_{mn}\left.{\bf w}({\bf x}_{mi}\!-\!{\bf x}_{mj})\right|_{i\neq j}.
\eea
The inverse matrix $G^{-1}_{AB}$ in \ur{m3} is understood according to the relation
$G^{-1}_{AC}G_{CB}=\delta_{AB}=\delta_{mn}\delta_{ij}$.

Note that the Coulomb bonds in \eq{G3} for the same-kind dyons have an opposite
sign from those for neighbor kind, and have a twice larger coefficient. The
coefficients, $-1,2,-1$, are actually the scalar products of simple roots of
the $SU(N)$ group, supplemented by an additional non-simple root to make the
matrix cyclic-symmetric. This remark allows the generalization of \eqs{G3}{W3}
to any Lie group.

The constructed $4KN\times 4KN$ metric tensor $g_{A\alpha,B\beta}$ is hyper-K\"ahler.
It means that there exist three ``complex structures'' $I(a),\,a=1,2,3,$
(all three are $4KN\times 4KN$ matrices) such that
\beq
I(a)g=gI(a)^T\quad({\rm ``T"\;means\;transposed})
\la{Icom}\eeq
and which satisfy the Pauli algebra,
\beq
I(a)I(b)=\epsilon^{abc}I(c)-\delta^{ab}{\bf 1}.
\la{Ialg}\eeq
Related to $I(a)$, there are three K\"ahler symplectic 2-forms
\beq
\omega(a)=\Omega(a)_{B\beta,C\gamma}\,dy_B^\beta\wedge dy_C^\gamma,
\quad \Omega(a)=-\Omega(a)^T,
\la{omega}\eeq
where
\beq
\Omega(a)=I(a)g.
\la{Omega}\eeq
The 2-forms $\omega(a)$ are closed:
\beq
d\omega(a)=0\quad{\rm or}\quad
\frac{\partial}{\partial y_A^\alpha}\Omega(a)_{B\beta,C\gamma}\,dy_A^\alpha
\wedge dy_B^\beta\wedge dy_C^\gamma=0.
\la{closure}\eeq
Explicitly, the three K\"ahler forms $\omega(a)$ have the same form as in Ref.~\cite{GM2}
for same-kind dyons, only $G_{AB}$ and ${\bf W}_{AB}$ should be now taken from
\eqs{G3}{W3}:
\beq
\omega(a)=2(d\psi_A+{\bf W}_{AA'}\!\cdot\!d{\bf x}_{A'})\wedge dx^a_{A}
-G_{BC}\,\epsilon^{abc}\,dx_B^b\wedge dx_C^c.
\la{omegaexplicit}\eeq

With $G_{AB}$ and ${\bf W}_{AB}$ given by \eqs{G3}{W3}, the three K\"ahler
forms $\omega(a)$ (or $\Omega(a)_{B\beta,C\gamma}$) are fixed from \eq{omegaexplicit},
and the complex structures $I(a)$ are found from inverting \eq{Omega}. We have checked
that the algebra \urs{Icom}{Ialg} is then satisfied for any choice of
${\bf w}({\bf x})$ in \eq{W3}. It is the closure of the two-forms, \eq{closure}, that
requests that ${\bf w}({\bf x})$ is the electric charge--magnetic charge interaction
potential satisfying the equation $\epsilon^{abc}\partial_b{\bf w}_c
=-{\bf x}_a/|{\bf x}|^3$.\\

We note further properties of the constructed $G_{AB}$ and ${\bf W}_{AB}$:
\begin{itemize}
\item symmetry: $G_{AB}=G_{BA},\,{\bf W}_{AB}={\bf W}_{BA}$, meaning, of course,
$G_{mi,nj}=G_{nj,mi},\,{\bf W}_{mi,nj}={\bf W}_{nj,mi}$
\item overall ``neutrality'':
$\sum_{nj}G_{mi,nj}=4\pi\nu_m,\,\sum_{mi}G_{mi,nj}=4\pi\nu_n,\,
\sum_{nj}{\bf W}_{mi,nj}=0,\,\sum_{mi}{\bf W}_{mi,nj}=0$
\item identity loss: dyons of the same kind are indistinguishable, meaning
mathematically that $\det G$ is symmetric under permutation of any pair of dyons
$(i\!\leftrightarrow\!j)$ of the same kind $m$
\item factorization: in the geometry when dyons fall into $K$ well separated
clusters of $N$ dyons of all kinds in each, $\det G$ factorizes
into a product of exact integration measures for $K$ KvBLL instantons,
$\det G=(\det G_1)^K$ where $G_1$ is given by \eq{G1}.
\end{itemize}
The integration measure over the moduli space of $K$ KvBLL instantons
of the $SU(N)$ gauge group is
\beq
\prod_{i=1}^K\prod_{m=1}^N \int d{\bf x}_{mi}\,d\psi_{mi}\,\sqrt{\det g},\qquad
\sqrt{\det g}=\det G.
\la{me3}\eeq
In deriving the last relation we notice that in the determinant, $d\psi_{mi}$
can be shifted by ${\bf W}_{mi,m'i'}\!\cdot\!d{\bf x}_{m'i'}$, hence
$\det g=(\det G)^3\,\det G^{-1}= (\det G)^2$, therefore $\sqrt{\det g}=\det G$
where the $KN\times KN$ matrix $G$ is given by \eq{G3}. We have also checked this result
by an explicit calculation of the determinant of the full $4KN\!\times\!4KN$
metric tensor $g$. Since $G$ is independent of the $U(1)$ angles $\psi_{mi}$,
integration over $\psi$ can be omitted.

\subsection{Dyons' fugacity}

Fugacity is a term from statistical mechanics of grand canonical ensembles (where
the number of particles is not fixed) denoting the weight with which a particle
contributes to the grand partition function. Let there be $K_m$ dyons of the $m^{\rm th}$
kind, $m=1...N$. For a neutral system of $K$ KvBLL instantons the number of dyons of
every kind is equal, $K_1=...=K_N=K$, however we shall consider the general case
of non-equal $K$'s for the time being: one can always project to the neutrality condition.
For an arbitrary set of $K$'s, $G$ is a $(K_1+...+K_N)\times (K_1+...+K_N)$ matrix given by \eq{G3}.

We write the partition function of the grand canonical ensemble as a sum over all
numbers of dyons of each kind:
\beq
{\cal Z}=\sum_{K_1...K_N}\frac{1}{K_1!...K_N!}\prod_{m=1}^N\,\prod_{i=1}^{K_m}
\int (d^3{\bf x}_{mi}\,f)\,\det G({\bf x}),
\la{Z1}\eeq
where $f$ is the $x$-independent factor -- the fugacity -- accompanying every
integral over ${\bf x}$. Since $\det G({\bf x})$ is symmetric under permutation
of same-kind dyons, the identity factorials are needed to avoid
counting the same configuration more than once. If one likes to impose the
overall neutrality condition, {\it viz.} that only configurations with the
equal number of dyons of different kind contribute to the partition function
($K_1=...=K_N=K$) one integrates \eq{Z1} over auxiliary angles:
\bea\la{Zneutr}
{\cal Z}^{\rm neutr}&=&\sum_{K_1...K_N}\int_0^{2\pi}
\frac{d\theta_1}{2\pi}...\frac{d\theta_N}{2\pi}\,
\frac{e^{i\theta_1(K_2-K_1)}}{K_1!}\ldots\frac{e^{i\theta_N(K_1-K_N)}}{K_N!}\,
\prod_{m=1}^N\,\prod_{i=1}^{K_m}\int (d^3{\bf x}_{mi}\,f)\,\det G({\bf x})\\
\n
&=&\sum_K\frac{1}{(K!)^N}\prod_{m=1}^N\,\prod_{i=1}^{K}
\int\!(d^3{\bf x}_{mi}\,f)\,\det G({\bf x}).
\eea
We shall see below, however, that the neutrality condition will be taken care
of dynamically, therefore the additional integration \ur{Zneutr} is in fact
unnecessary.

As fugacity is $x$-independent it can be established from the limit when all dyons are
grouped into $N$-plets of different-kind dyons, forming infinitely dilute neutral KvBLL
instantons, such that the measure factorizes into a product of individual instanton measures.
The latter is known to be~\cite{DG05}
\beq
\prod_{m=1}^N \int\!d^3{\bf x}_m\,\det G_1({\bf x})\,2^{2N}\pi^{3N}
\la{1inst1}\eeq
where $G_1({\bf x})$ is the $N\times N$ matrix \ur{G1} for {\em one} KvBLL instanton, see \eq{me1}.
This must be multiplied by the factor $\left[(\mu^4/T)\,/\sqrt{2\pi g^2}^{\;4}\right]^N$
coming from $4N$ zero modes of the instanton. Here $\mu$ is the ultraviolet
cutoff and $g$ is the bare coupling constant given at that cutoff~\cite{DG05,DGPS}.
Multiplication by this factor makes \eq{1inst1} dimensionless, as it should be.
In addition, \eq{1inst1} is multiplied by the exponent of minus the classical action of the
instanton, equal to $(\Lambda/\mu)^{\frac{11}{3}N}$ where $\Lambda$ is
the Yang--Mills scale parameterizing the coupling constant in the Pauli--Villars regularization
scheme, and by the dimensionless factor $\left[\det(-D^2)\right]^{-1}$ where $D^2$
is the Laplace operator in the instanton background. The last factor arises from
integration over non-zero modes; it is understood that the small-oscillation
determinant is normalized to the free (zero field) determinant and UV-regularized
by the Pauli--Villars method. It is known that the normalized and regularized
$\left[\det(-D^2)\right]^{-1}$ is proportional to $\mu^{-\frac{N}{3}}$, times the
exponent of minus the perturbative potential energy \ur{Ppert}, times a slowly
varying function of dyon separations~\cite{DGPS,Gromov}.

Combining all factors we observe that the Pauli--Villars mass $\mu$ cancels out
(as it should be in a renormalizable theory) and we obtain the dyon fugacity
\beq
f=\frac{\Lambda^4}{T}\,\frac{4\pi}{g^4}\,c
\la{f1}\eeq
where $c$ is proportional to $\Lambda^{-\frac{1}{3}}$; it is made dimensionless by
a combination of temperature and dyon separations. The relative (un)importance of
$c$ in the dynamics of the ensemble, as compared to the measure factor
\ur{Z1}, is illustrated by the powers of $\Lambda$: their ratio is
$(\!-\!1/3):4=\!-\!1/12$. For the time being we shall set $c\!=\!1$ and recall it
in the discussion in Section IV. The coupling $g^2$ in\ur{f1}
starts to `run' at the two-loop level not included here. Ultimately, its
precise argument is determined self-consistently from the action density of the
ensemble~\cite{DP1}. In the study of the large-$N$ behavior it will be important
that $c={\cal O}(1)$ whereas $1/g^4={\cal O}(N^2)$, hence the fugacity $f={\cal O}(N^2)$.

\section{Dyon partition function as a quantum field theory}

We now face an interesting problem of finding the correlation functions in the
ensemble of dyons whose grand partition function is given by \eq{Z1}. The
renormalized Yang--Mills scale parameter $\Lambda$ creeps in via the
fugacity \ur{f1}, therefore all physical quantities will be henceforth expressed
through $\Lambda$. The temperature also enters explicitly via \eq{f1}; the
temperature factors are understood in all Coulomb bonds in the matrix $G$
\ur{G3} as well, to make them dimensionless. Thus, the partition function and
the ensuing correlation functions depend, generally, both on $\Lambda$ and $T$.

The ensemble defined by the partition function \ur{Z1} is a very unusual one,
as it is governed by the determinant of a matrix $G$ whose dimension is equal to
the number of particles, and not by the exponent of the interaction energy, as
is common in statistical mechanics. Of course, one can always write
$\det G=\exp\Tr\log G\equiv \exp(U_{\rm int})$, however then the interaction
potential $U_{\rm int}$ will contain not only 2-body, but also 3,4,5... -body
forces that are increasingly important. At the same time, the statistical mechanics
of an ensemble governed by the determinant-induced interactions can be transformed
into an equivalent quantum field theory which considerably simplifies its handling.

To that end, we first notice that a matrix determinant can be presented as a result
of the integration over a finite number of anticommuting Grassmann variables~\cite{Berezin},
\beq
\det
G=\int\!\prod_A d\psi_A^\dagger\,d\psi_A\,
\exp\left(\psi_A^\dagger\,G_{AB}\,\psi_B\right),
\la{detG1}\eeq
where the usual convention~\cite{Berezin} for anticommuting integration variables
is understood:
\bea\n
&&\psi_A\psi_B+\psi_B\psi_A=0,\quad\psi_A^\dagger\psi_B^\dagger+\psi_B^\dagger\psi_A^\dagger=0,
\quad\psi_A^\dagger\psi_B+\psi_B\psi_A^\dagger=0\quad {\rm for\;any}\;A,B,\\
\n\\
&&\int\!d\psi_A^\dagger\,d\psi_A=0,\qquad
\int\!d\psi_A^\dagger\,d\psi_A\,\psi_A^\dagger\,\psi_A=1.
\la{Grass}\eea
In our case $A=(mi)$ is a multi-index where $m=1...N$ is the dyon kind,
and $i=1...K_m$ is the number of a dyon of the $m^{\rm th}$ kind. We rewrite
identically the partition function \ur{Z1} as
\beq
{\cal Z}=\sum_{K_1...K_N}\frac{f^{K_1} ... f^{K_N}}{K_1!...K_N!}
\prod_{m=1}^N\,\prod_{i=1}^{K_m}\int\!d^3{\bf x}_{mi}\int\!d\psi_{mi}^\dagger\,
d\psi_{mi}\,\exp\left(\psi_{mi}^\dagger\,G_{mi,nj}\,\psi_{nj}\right)
\la{Z2}\eeq
where $G_{mi,nj}$ is a matrix made of Coulomb interactions, \eq{G3}, and $f$ is the
fugacity \ur{f1} where we put $c\!=\!1$. Having obtained $G$ in the exponent,
it is now possible to express its Coulomb matrix elements from path integrals,
extending the Polyakov trick~\cite{Polyakov77} to anticommuting variables.

\subsection{Off-diagonal elements: ghost fields}

We first present the off-diagonal ($i\neq j$) elements of $\psi^\dagger G\psi$ by means
of a functional integration over anticommuting (or ghost) fields.
In the next subsection we present the diagonal ($i=j$) elements with the help of
a functional integration over commuting (boson) fields.

Let us consider the Gaussian path integral over $N$ anticommuting fields $\chi_m({\bf x})$
coupled to the anticommuting source $\sum_i\left[\psi_{mi}\delta({\bf x}\!-\!{\bf
x}_{mi})-\psi_{m\!-\!1,i}\delta({\bf x}\!-\!{\bf x}_{m\!-\!1,i})\right]$:
\bea\la{Y1}
{\cal Y}&=&\prod_{m=1}^N\int\!D\chi_m^\dagger D\chi_m\,\exp\int\!d{\bf x}
\sum_m\left\{\frac{T}{4\pi}\,
{\boldmath \partial}\chi_m^\dagger\!\cdot\!{\boldmath \partial}\chi_m \right. \\
\n
&+&\left.i\sum_i\left[\left(\psi_{mi}^\dagger\chi_m({\bf x})
+\chi_m^\dagger({\bf x})\psi_{mi}\right)\delta({\bf x}\!-\!{\bf x}_{mi})
-\left(\psi_{m\!-\!1,i}^\dagger\chi_m({\bf x})
+\chi_m^\dagger({\bf x})\psi_{m\!-\!1,i}\right)
\delta({\bf x}\!-\!{\bf x}_{m\!-\!1,i})\right]\right\}.
\eea
Although we do not write it explicitly to save space, we assume that ${\cal Y}$
is normalized to the same path integral with the kinetic term but without the
source term. The subscript $m$ is periodic: $m=N+1$ is equivalent to $m=1$
and $m=0$ means $m=N$.

The path integration of an action that is quadratic in anticommuting variables
is performed in the same way as the Gaussian path integral over bosonic
variables, with the result
\beq
{\cal Y}=\exp\left[\frac{1}{T}\sum_{m,\,i,j}
\left(\frac{\psi_{mi}^\dagger\psi_{mj}}{|{\bf x}_{mi}\!-\!{\bf x}_{mj}|}
-\frac{\psi_{mi}^\dagger\psi_{m\!-\!1,j}}{|{\bf x}_{mi}\!-\!{\bf x}_{m\!-\!1,j}|}
-\frac{\psi_{m\!-\!1,i}^\dagger\psi_{mj}}{|{\bf x}_{m\!-\!1,i}\!-\!{\bf x}_{mj}|}
+\frac{\psi_{m\!-\!1,i}^\dagger\psi_{m\!-\!1,j}}{|{\bf x}_{m\!-\!1,i}\!-\!{\bf x}_{m\!-\!1,j}|}
\right)\right].
\la{Y2}\eeq
Owing to the cyclic summation over $m$, the last term doubles the first one, and
we correctly reproduce the off-diagonal ($i\neq j$) part of
$\psi_{mi}^\dagger G_{mi,nj}\psi_{nj}$ in \eq{Z2} (cf. the second line in
\eq{G3}). However, the sum in \eq{Y2} contains an extra diagonal divergent term
$2\psi_{mi}^\dagger\psi_{mi}/|{\bf x}_{mi}\!-\!{\bf x}_{mi}|$ which is absent
in \eq{Z2} and hence should be canceled.

\subsection{Diagonal elements: boson fields}

Next, we present the diagonal ($i\!=\!j$) part of $\psi^\dagger G\psi$ by means
of a Gaussian integration over bosonic fields $v_m,w_m$.

Let us consider
\bea\la{X1}
{\cal X}&=&\prod_{m=1}^N\int\!Dv_m Dw_m\,\exp\int\!d{\bf x}
\sum_m\left\{\frac{T}{4\pi}\,
{\boldmath \partial}v_m\!\cdot\!{\boldmath \partial}w_m \right. \\
\n
&+&\!\!\!\!\left.\sum_i\left[\left(\psi_{mi}^\dagger\psi_{mi}
\delta({\bf x}\!-\!{\bf x}_{mi})-\psi_{m\!-\!1,i}^\dagger\psi_{m\!-\!1,i}
\delta({\bf x}\!-\!{\bf x}_{m\!-\!1,i})\right)v_m({\bf x})
+\left(\delta({\bf x}\!-\!{\bf x}_{mi})-\delta({\bf x}\!-\!{\bf x}_{m\!-\!1,i})\right)
w_m({\bf x})\right]\right\}.
\eea
To make this path integral formally convergent one assumes that the integration
over either $v_m$ or $w_m$ goes along the imaginary axis. As in the case of
ghost fields in the previous subsection, we do not write it explicitly but assume that
${\cal X}$ is normalized to the same path integral with the kinetic term but
without the source term (the second line in \eq{X1}).

Integrating \ur{X1} over $w_m$ we obtain a functional $\delta$-function:
$$
\delta\left(-\frac{T}{4\pi}\,\partial^2v_m+\sum_i
\left[\delta({\bf x}\!-\!{\bf x}_{mi})-\delta({\bf x}\!-\!{\bf x}_{m\!-\!1,i})
\right]\right),
$$
whose solution is
\beq
v_m({\bf x})=-\frac{1}{T}\,\sum_i\left(\frac{1}{|{\bf x}\!-\!{\bf x}_{mi}|}
-\frac{1}{|{\bf x}\!-\!{\bf x}_{m\!-\!1,i}|}\right),\qquad \sum_m v_m({\bf x})=0.
\la{vm}\eeq
The Jacobian following from the $\delta$-function,
$\det\left(-(T/4\pi)\partial^2\right)$, cancels with the same Jacobian from the
normalization integral. Substituting $v_m(x)$ from \eq{vm} back into \eq{X1} and
using the cyclic symmetry of the summation over $m$ we obtain
\beq
{\cal X}=\exp\left[-\frac{1}{T}\sum_{m,\,i,j}
\left(2\frac{\psi_{mi}^\dagger\psi_{mj}}{|{\bf x}_{mi}\!-\!{\bf x}_{mj}|}
-\frac{\psi_{mi}^\dagger\psi_{mi}}{|{\bf x}_{mi}\!-\!{\bf x}_{m\!-\!1,j}|}
-\frac{\psi_{mi}^\dagger\psi_{mi}}{|{\bf x}_{mi}\!-\!{\bf x}_{m\!+\!1,j}|}
\right)\right].
\la{X2}\eeq
The divergent term at $i\!=\!j$, namely
$-2\psi_{mi}^\dagger\psi_{mi}/|{\bf x}_{mi}\!-\!{\bf x}_{mi}|$,
cancels exactly the unwanted extra term in \eq{Y2}, and we reproduce precisely
the diagonal term in $\psi_{mi}^\dagger G_{mi,nj}\psi_{nj}$ of \eq{Z2} (cf. the first line in
\eq{G3}).

Thus, we have fully reproduced the factor $\exp(\psi^\dagger G\psi)$ in the
partition function \ur{Z2} with the help of the integration over anticommuting ghost
($\chi^\dagger_m,\chi_m$) and ordinary boson ($v_m,w_m$) variables. The Coulomb
interactions have been traded for kinetic energy terms of those fields.
Apparently, $v_m$ are $U(1)^{N\!-\!1}$ Abelian electric potentials,
and $w_m$ are their dual fields.

\subsection{Synthesis: the equivalent quantum field theory}

We now use \eqs{Y1}{X1} to rewrite identically the partition function \ur{Z2}.
We have
\bea\la{Z3}
{\cal Z}&=&\prod_{m=1}^N\int\!D\chi_m^\dagger\,D\chi_m\,Dv_m\,Dw_m\,\exp\int\!d{\bf x}
\sum_m\left[\frac{T}{4\pi}\,\left({\boldmath \partial}\chi_m^\dagger\!\cdot\!{\boldmath
\partial}\chi_m+{\boldmath \partial}v_m\!\cdot\!{\boldmath\partial}w_m\right)\right]\\
\n
&\times &\sum_{K_1=0}^\infty\frac{f^{K_1}}{K_1!}\!\left(\int\!d{\bf x}_1\!\int\!d\psi_1^\dagger
d\psi_1\,\exp\left[4\pi\nu_1\psi_1^\dagger\psi_1
+i\psi_1^\dagger\left(\chi_1(x_1)\!-\!\chi_2(x_1)\right)
+i\left(\chi_1^\dagger(x_1)\!-\!\chi_2^\dagger(x_1)\right)\psi_1\right.\right.\\
\n
&&\hskip 5.5true cm +\left.\left.\left(v_1(x_1)\!-\!v_2(x_1)\right)\psi_1^\dagger\psi_1
+\left(w_1(x_1)\!-\!w_2(x_1)\right)\right]\;\right)^{K_1}\\
\n\\
\n
&\times &\ldots \\
\n\\
\n
&\times &\!\!\!\!\sum_{K_N=0}^\infty\frac{f^{K_N}}{K_N!}\!\left(\!\int\!d{\bf x}_N\!\!
\int\!d\psi_N^\dagger d\psi_N\,\exp\left[4\pi\nu_N\psi_N^\dagger\psi_N
+i\psi_N^\dagger\left(\chi_N(x_N)\!-\!\chi_1(x_N)\right)
+i\left(\chi_N^\dagger(x_N)\!-\!\chi_1^\dagger(x_N)\right)\psi_N\right.\right.\\
\n
&&\hskip 5.5true cm +\left.\left.\left(v_N(x_N)\!-\!v_1(x_N)\right)\psi_N^\dagger\psi_N
+\left(w_N(x_N)\!-\!w_1(x_N)\right)\right]\;\right)^{K_N}.
\eea
In writing \eq{Z3} we have used that $K_m$ identical integrals over
$d{\bf x}_{mi}\,d\psi_{mi}^\dagger d\psi_{mi}$ appear in the
partition function, where $i\!=\!1...K_m$ is a `dumb' index labeling integration
variables. Therefore one representative of such integral for every dyon kind $m$
is taken to the power $K_m$.

In each line in \eq{Z3} integration over $d\psi^\dagger_m d\psi_m$ can be
trivially performed, given the rules \ur{Grass}: it reduces to expanding the
exponents in \eq{Z3} to the terms bilinear in $\psi^\dagger_m,\psi_m$. For
example, for $m=1$ we get
\bea\n
&&\sum_{K_1=0}^\infty\frac{1}{K_1!}\left(f\!\int\!d{\bf x}\,
\left[4\pi\nu_1+(\chi^\dagger_1\!-\!\chi^\dagger_2)(\chi_1\!-\!\chi_2)({\bf x})
+(v_1-v_2)({\bf x})\right]e^{(w_1-w_2)({\bf x})}\right)^{K_1}\\
&&=\exp\left(f\!\int\!d{\bf x}\,\left[4\pi\nu_1+(\chi^\dagger_1\!-\!\chi^\dagger_2)
(\chi_1\!-\!\chi_2)+(v_1-v_2)\right]e^{w_1-w_2}\right)
\eea
and similarly for other values of $m$. We obtain
\bea\la{Z4}
{\cal Z}&=&\prod_{m=1}^N\int\!D\chi_m^\dagger\,D\chi_m\,Dv_m\,Dw_m\,\exp\int\!d{\bf x}
\sum_m\left\{\frac{T}{4\pi}\,\left({\boldmath \partial}\chi_m^\dagger\!\cdot\!{\boldmath
\partial}\chi_m+{\boldmath \partial}v_m\!\cdot\!{\boldmath\partial}w_m\right)\right.\\
\n
&+&\left.f\left[4\pi\nu_m+(v_m-v_{m\!+\!1})
+(\chi^\dagger_m-\chi^\dagger_{m\!+\!1})(\chi_m-\chi_{m\!+\!1})\right]
e^{w_m-w_{m\!+\!1}}\right\}.
\eea
Given the cyclic symmetry in the summation over $m$, the last line can be
rewritten in a more nice way. We introduce the function
\beq
{\cal F}(w)\equiv\sum_{m\!=\!1}^N e^{w_m-w_{m\!+\!1}}
\la{calF}\eeq
and recall that $\nu_m=\mu_{m\!+\!1}-\mu_m$ where $\mu_m$ are the eigenvalues
of the Polyakov line, see the Introduction. The second line in \eq{Z3} can be
written as
$$
f\left[(-4\pi\mu_m+v_m)\frac{\partial{\cal F}}{\partial w_m}
+\chi^\dagger_m\,\frac{\partial^2{\cal F}}{\partial w_m\partial
w_n}\,\chi_n\right]\,,
$$
(summation over repeated indices is understood) where
\bea\n
\frac{\partial{\cal F}}{\partial w_m}&=&e^{w_m-w_{m+1}}-e^{w_{m-1}-w_m},\\
\n
\frac{\partial^2{\cal F}}{\partial w_m\partial w_n}&=&
\delta_{mn}\,\left(e^{w_m-w_{m+1}}+e^{w_{m-1}-w_m}\right)
-\delta_{m,n-1}e^{w_m-w_{m+1}}-\delta_{m,n+1}e^{w_{m-1}-w_m}\,.
\eea

The final result for the dyon partition function is
\bea\la{Z5}
{\cal Z}&=&\int\!D\chi^\dagger\,D\chi\,Dv\,Dw\,\exp\int\!d^3x
\left\{\frac{T}{4\pi}\,\left(\partial_i\chi_m^\dagger\partial_i\chi_m
+\partial_iv_m\partial_iw_m\right)\right.\\
\n
&+&\left.f\left[(-4\pi\mu_m+v_m)\frac{\partial{\cal F}}{\partial w_m}
+\chi^\dagger_m\,\frac{\partial^2{\cal F}}{\partial w_m\partial w_n}\,\chi_n\right]
\right\},
\eea
\Eq{Z5} should be divided by the normalization integral being the same expression but
with zero fugacity $f$. In fact the normalization integral is unity and can be omitted.
Indeed, integrating over $v_m$ gives $\delta(-(T/4\pi)\partial^2w_m)$ whose only
solution is $w_m={\rm const}$, whereas the Jacobian is ${\rm det}^{-1}(-(T/4\pi)\partial^2)$.
This Jacobian, however, is immediately canceled by the integral over the ghost
fields $\chi_m$. Therefore, the quantum field theory defined by \eq{Z5}
is the full result for the dyon partition function.

\section{Ground state: `confining' holonomy preferred}


The fields $v_m$ enter the partition function \ur{Z5} only linearly. Therefore,
they can be integrated out right away, giving rise to a $\delta$-function
\beq
\int\!Dv_m\quad\longrightarrow\quad \delta\left(-\frac{T}{4\pi}\partial^2w_m
+f\frac{\partial{\cal F}}{\partial w_m}\right).
\la{UD}\eeq
This $\delta$-function restricts possible fields $w_m$ over which one still has
to integrate in \eq{Z5}. Let $\bar w_m$ be a solution to the argument of the
$\delta$-function. Integrating over small fluctuations about $\bar w$ gives
the Jacobian
\beq\la{Jac}
{\rm Jac}={\rm det}^{-1}\left(-\frac{T}{4\pi}\partial^2\delta_{mn}
+\left.f\frac{\partial^2{\cal F}}{\partial w_m\partial w_n}\,
\right|_{w=\bar w}\,\right)\,.
\eeq
Remarkably, exactly the same functional determinant (but in the numerator)
arises from integrating over the ghost fields, in the same background $\bar w$:
\beq
\int\!D\chi^\dagger D\chi\exp\int\!d^3x
\left[\frac{T}{4\pi}\,\partial_i\chi_m^\dagger\partial_i\chi_m
+f\chi^\dagger_m\,\frac{\partial^2{\cal F}}{\partial w_m\partial w_n}\,\chi_n\right]
=\det\left(-\frac{T}{4\pi}\partial^2\delta_{mn}+f\frac{\partial^2{\cal F}}
{\partial w_m\partial w_n}\right).
\la{detchi}\eeq
Therefore, all quantum corrections cancel {\em exactly} between the boson and
ghost fields (a characteristic feature of supersymmetry), and the ensemble of dyons
is basically governed by a classical field theory~\cite{footnote15}.

To find the ground state we examine the fields' potential energy being
$-4\pi f\mu_m\partial{\cal F}/\partial w_m$ which we prefer to write restoring
$\nu_m=\mu_{m\!+\!1}-\mu_m$ and ${\cal F}$ as
\beq
{\cal P}=4\pi f\sum_m \nu_m\,e^{w_m-w_{m\!+\!1}}.
\la{calP1}\eeq
For constant fields $w_m$, this is multiplied by the volume, therefore one has to find the
stationary point for any given set of $\nu_m$'s. It leads to the equations
\beq
\frac{\partial {\cal P}}{\partial w_1}=4\pi f\left(\nu_1e^{w_1-w_2}-\nu_Ne^{w_N-w_1}\right)=0,\qquad
\frac{\partial {\cal P}}{\partial w_2}=4\pi f\left(\nu_2e^{w_2-w_3}-\nu_1e^{w_1-w_2}\right)=0,
\quad\ldots\quad
\la{extr1}\eeq
whose solution is
\beq
e^{w_1-w_2}=\frac{(\nu_1\nu_2\nu_3...\nu_N)^{\frac{1}{N}}}{\nu_1},\quad
e^{w_2-w_3}=\frac{(\nu_1\nu_2\nu_3...\nu_N)^{\frac{1}{N}}}{\nu_2},\quad
{\rm etc.}
\la{extr2}\eeq
The solution corresponds to all terms in \eq{calP1} being equal, despite {\it a priori}
non-equal $\nu_m$'s. Putting it back into \eq{calP1} we obtain
\beq
{\cal P}=4\pi f N (\nu_1\nu_2...\nu_N)^{\frac{1}{N}},\qquad
\nu_1+\nu_2+...+\nu_N=1.
\la{calP2}\eeq
The maximum is achieved when all $\nu$'s are equal:
\beq
\nu_1=\nu_2=...=\nu_N=\frac{1}{N},\qquad {\cal P}^{\rm max}=4\pi f.
\la{numax}\eeq
Equal $\nu$'s correspond to the ``maximal non-trivial'' or ``confining''
holonomy, see \eq{muconf}. Since there are no quantum corrections, the
free energy of the dyons ensemble is simply proportional to the classical
potential energy, $F=-{\cal P}V$. Therefore, the maximum of ${\cal P}$
corresponds to the minimum of the free energy. Thus the free energy of the
grand canonical ensemble has the minimum at the confining values of the holonomy
(see the Introduction). In the minimum the free energy is
\beq
F^{\rm min}=-4\pi f V=-\frac{16\pi^2}{g^4}\,\Lambda^4\,\frac{V}{T}\,=\,
\frac{N^2}{4\pi^2}\,\frac{\Lambda^4}{\lambda^2}\,\frac{V}{T}\,,
\qquad\quad \lambda\equiv\frac{\alpha_s N}{2\pi}=
\frac{g^2N}{8\pi^2},
\la{freeen}\eeq
and there are no corrections to this result. In the last equation we have introduced
the $N$-independent 't Hooft coupling $\lambda$.

Let us make a few comments. First, the free energy \ur{freeen} has the correct behavior
at large $N$. Second, $V/T=V^{(4)}$ is in fact the $4d$ volume of the $R^3\times S^1$ space.
Although we do not expect our theory to be valid at small temperatures (where the measure
we use for same-kind dyons is probably incomplete), \eq{freeen} can be formally
extended to the zero-temperature limit, as it correctly reproduces the extensive dependence
on the $4d$ volume. Third, \eq{freeen} gives in fact the density of dyons. One can introduce
separate fugacities $f_m$ for dyons of the $m^{\rm th}$ kind into the partition
function \ur{Z2}; then the average number of dyons is found from the obvious
relation
$$<\!K_m\!>=\left.\frac{\partial \log{\cal Z}}{\partial \log f_m}\right|_{f_m=f}.$$
With separate fugacities, the result \ur{freeen} is modified by replacing
$f\to(f_1f_2...f_N)^{1/N}$, hence
\beq
<\!K_m\!>=-\left.f_m\frac{\partial F^{\rm min}}{\partial f_m}\right|_{f_m=f}
=\frac{1}{N}\,4\pi f\,V=\frac{N}{4\pi^2}\,\frac{\Lambda^4}{\lambda^2}\,V^{(4)},
\la{avn}\eeq
{\it i.e.} a finite (and equal) density of each kind of dyons in the 4-volume,
meaning also the finite density of the KvBLL instantons. From the 3-dimensional
point of view, the $3d$ density of dyons (and KvBLL instantons) is increasing
as the temperature goes down: there are more and more instantons sitting on top
of each other in $3d$ but spread over the compactified time direction.

Let us add a few comments of speculative nature as they extend what is actually
done here. We attempt to make contact with the phenomenology of the pure
glue $SU(N)$ Yang--Mills theory. In the real world there must be as many
anti-self-dual dyons in the vacuum as there are dual ones, up to thermodynamic
fluctuations $\sim \sqrt{V}$. For a crude estimate, we make the simplest
assumption that adding anti-self-dual dyons just doubles the free energy.
If the topological angle $\theta$ is introduced, one has to change dyon fugacities
$f\to f\,e^{i\theta/N}$ and anti-dyon fugacities $f\to f\,e^{-i\theta/N}$,
such that the KvBLL instanton whose fugacity is $f^N$ acquires a phase $e^{i\theta}$
and the anti-instanton acquires a phase $e^{-i\theta}$~\cite{Zhitnitsky}.
After minimization in $w_m$ and $\nu_m$ which goes as before, the free
energy
\ur{freeen} becomes
\beq
F=-4\pi f\,2\cos\frac{\theta}{N}\,V
=-\frac{16\pi^2}{g^4}\,\Lambda^4\,2\cos\frac{\theta}{N}\,\frac{V}{T}
\la{theta}\eeq
leading to the topological susceptibility
\beq
<\!Q_T^2\!>=\int\!d^4x\,\left<\!\frac{\Tr F\tilde F(x)}{16\pi^2}
\frac{\Tr F\tilde F(0)}{16\pi^2}\!\right>
=\frac{1}{V^{(4)}}\left.\frac{\partial^2F}{\partial\theta^2}\right|_{\theta=0}
=\frac{32\pi^2}{N^2g^4}\,\Lambda^4
=\frac{1}{2\pi^2}\frac{\Lambda^4}{\lambda^2}\,.
\la{topsusc}\eeq
We see that the topological susceptibility is stable at large $N$ as it is
expected from the $N$-counting rules.

The free energy is related, via the trace anomaly, to the so-called gluon
condensate~\cite{DP1,D02}
$$F\simeq-\frac{11N}{12}\,\frac{<\!\Tr F_{\mu\nu}^2\!>}{16\pi^2}\,V^{(4)}$$
from where we find
\beq
\frac{<\!\Tr F_{\mu\nu}^2\!>}{16\pi^2}\simeq
N\,\frac{12}{11}\,\frac{1}{2\pi^2}\,\frac{\Lambda^4}{\lambda^2}
=N\,\frac{12}{11}\,<\!Q_T^2\!>\,.
\la{gluoncond}\eeq
It is the expected $N$-dependence of the condensate.

As the temperature increases, the perturbative potential energy
\ur{Ppert} becomes increasingly important since its contribution grows as $T^4$
with respect to the non-perturbative one. The perturbative energy arises
from the small-oscillation determinant $\left[\det(-D^2)\right]^{-1}$ denoted
as $c$ in \eq{f1}. If we naively add up the dyon-induced
free energy \ur{freeen} and the perturbative energy \ur{Ppertmax} both computed at
the maximally non-trivial holonomy \ur{numax}, we obtain the full free energy
$$\left(-\frac{32\pi^2}{g^4}\,\frac{\Lambda^4}{T}
+T^3\,\frac{(2\pi)^2}{180}\,\frac{N^4-1}{N^2}\right)V\,.$$
It becomes positive and hence less favorable than the zero energy of the
trivial holonomy at the temperature
\beq
T_c^4=\frac{45}{2\pi^4}\,\frac{N^4}{N^4-1}\,
\frac{\Lambda^4}{\lambda^2}.
\la{Tcrit}\eeq
At this temperature, the deconfinement phase transition is expected.
We see that $T_c$ is stable in $N$ as it should be on general grounds.
For a numerical estimate at $N\!=\!3$, we take $\lambda=1/4$ compatible with
the commonly assumed freezing of $\alpha_s$ at the value of 0.5, and
$\Lambda=200\,{\rm MeV}$ in the Pauli--Villars scheme.
We then obtain from \eqsss{topsusc}{Tcrit} the topological susceptibility,
the gluon condensate and the critical temperature $(189\,{\rm MeV})^4,\;
(255\,{\rm MeV})^4$ and $278\,{\rm MeV}$, respectively, being in reasonable
agreement with the phenomenological and lattice values. More robust quantities
(both from the theoretical and lattice viewpoints) are those measured in units
of the string tension; such comparison will be made in the next Section.

From now on, we shall assume we are far enough below the critical temperature,
so that the minimum of the free energy implies the ``confining'' holonomy, \eq{numax}.
From \eq{extr2} we learn that at the minimum all constant parts of $w_m$'s are equal
(up to possible difference in $2\pi i k$ with integer $k$, which does not
change the exponents of $w$). Let us note that had we imposed the overall
neutrality condition of the dyon ensemble by an additional integration over the
$\theta$ angles (see \eq{Zneutr}) it would be equivalent to shifting $w_m\to
w_m+i\theta_m$. Since in \eq{Z5} one integrates over all functions $w_m$ including
their constant parts, an additional integration over $\theta$'s is unnecessary,
and the neutrality condition is imposed automatically.

The triviality of the free energy \ur{freeen} (which is due to the cancelation between
boson and ghost quantum determinants) does not mean the triviality of
the ensemble: dyons are in fact strongly correlated, as we shall see in the last sections.
To study correlations, one has to insert source terms into the partition function
\ur{Z5}. With the sources switched on, the fields $w_m$ are allowed to be
$x$-dependent. Therefore, one has to retain the term
$-4\pi f\mu_m\partial{\cal F}/\partial w_m$ which we rewrite using \eq{calP1}
and \eq{numax} as
\beq
{\rm Action}=\int\!d^3x\,\frac{4\pi f}{N}{\cal F}(w)
\la{action}\eeq
where ${\cal F}(w)$ is defined in \eq{calF}.

Finally, we note that the equation of motion for the fields $w_m$, following from
the $\delta$-function \ur{UD}, is known as the periodic Toda lattice~\cite{Toda}
which has plenty of soliton solutions. In particular, there are many one-dimensional
domain-wall solutions interpolating between $w_m\!-\!w_n=2\pi i k_{mn}$ and
$2\pi i k'_{mn}$ where $k,k'$ are integers. Why do not they contribute to the partition function?
The answer is that any soliton is $x$-dependent, and an overall shift of the soliton
is a zero mode of the operator \ur{Jac} resulting in an integration over the soliton
position in space. However, it is also a zero mode of the identical operator for
ghosts \ur{detchi}, leading to a vanishing ghost determinant. Therefore, any soliton
gives a zero contribution to the partition function. However, solitons may and will
generally contribute to the correlation functions.

\section{Correlation function of Polyakov lines}

In the gauge where $A_4({\bf x})$ is chosen to be time-independent the Polyakov line
is $\Tr L({\bf z})=\Tr \exp(iA_4({\bf z})/T)$. The $A_4$ field of $K$ KvBLL instantons away
from their cores is Abelian~\cite{KvBSUN} and can be gauge-chosen to be
diagonal:
$$A_4({\bf z})/T=\delta_{mn}\left[2\pi \mu_m+\frac{1}{2T}\sum_i\left(
\frac{1}{|{\bf z}-{\bf x}_{mi}|}-\frac{1}{|{\bf z}-{\bf x}_{m\!-\!1,i}|}\right)\right].$$
Comparing it with \eq{vm} we observe that $A_4$ can be written as
\beq
A_4({\bf z})/T={\rm diag}\left(2\pi\mu_m-\half v_m({\bf z})\right),\qquad
\Tr L({\bf z})=\sum_m\exp\left(2\pi i\mu_m-\frac{i}{2}v_m({\bf z})\right).
\la{Av}\eeq
Therefore, to compute the vacuum average of any number of Polyakov lines, one
has to add a source term to the partition function \ur{Z5}:
\beq
\sum_{m_1,m_2...} \exp\left[\epsilon_{m_1}\!\!\left(2\pi i \mu_{m_1}\!
-\!\frac{i}{2}\!\int\!\!d{\bf x}\,v_m({\bf x})\delta_{mm_1}\delta({\bf x}\!-\!{\bf z_1})\right)
+\epsilon_{m_2}\!\!\left(2\pi i \mu_{m_2}\!-\!\frac{i}{2}\!\int\!\!d{\bf x}\,v_m({\bf x})
\delta_{mm_2}\delta({\bf x}\!-\!{\bf z_2})\right)+\ldots\right],
\la{source1}\eeq
where ${\bf z}_{1,2...}$ are the points in space where Polyakov lines are placed
and $\epsilon_{m_1,m_2...}=\pm 1$ depending on whether one takes
$L=\exp(iA_4/T)$ or $L^\dagger=\exp(-iA_4/T)$.

The source term is linear in $v_m$ which means that integration over $v_m$ in
the partition function with a source produces a $\delta$-function \ur{UD} as before whose
argument is now shifted by the source:
\beq
\int\!Dv_m\quad\longrightarrow \quad
\delta\left(-\frac{T}{4\pi}\partial^2w_m+f\frac{\partial{\cal F}}{\partial w_m}
-\epsilon_{m_1}\frac{i}{2}\,\delta({\bf x}\!-\!{\bf z_1})\delta_{mm_1}
-\epsilon_{m_2}\frac{i}{2}\,\delta({\bf x}\!-\!{\bf z_2})\delta_{mm_2}
-\ldots\right).
\la{UDPol}\eeq
The correlation function of any number of widely separated Polyakov lines in
the fundamental representation is given by the path integral with $\delta$-functions:
\bea\la{corr0}
&&\left<\Tr L({\bf z}_1)\Tr L({\bf z}_2)\ldots\right>
=\sum_{m_1,m_2,...}e^{2\pi i(\mu_{m_1}+\mu_{m_2}+\ldots)}\int\!Dw_m\exp\left(\!\int\!d{\bf x}\,
\frac{4\pi f}{N}{\cal F}(w)\right)\\
\n
&&\cdot\!\prod_m\delta\!\left(\!-\frac{T}{4\pi}\partial^2w_m\!
+\!f\frac{\partial{\cal F}}{\partial w_m}
\!-\!\frac{i}{2}\,\delta({\bf x}\!-\!{\bf z_1})\delta_{mm_1}
\!-\!\frac{i}{2}\,\delta({\bf x}\!-\!{\bf z_2})\delta_{mm_2}\!-\!\ldots\right)
\det\!\left(\!-\frac{T}{4\pi}\partial^2\delta_{mn}\!+\!
f\frac{\partial^2{\cal F}}{\partial w_m\partial w_n}\right).
\eea
It is understood that \eq{corr0} is divided by the same expression but without
the source. The last factor comes from integrating over the ghost fields.

The strategy is to find all possible solutions of the $\delta$-functions,
substitute them into the action \ur{action} and to sum over $m_{1,2...}$.
Note that, whatever functions $w_m$ solve the $\delta$-functions, the Jacobian
arising from those $\delta$-functions is again canceled exactly by the ghost
determinant. Therefore, there will be no corrections to a classical
calculation.

\subsection{Average of a single line}

The average $<\!\Tr L\!>$ is expected to be zero for the confining holonomy but let
us check how it follows from the general \eq{corr0}. In this case there is only
one $\delta$-function source in \eq{corr0}. One has to solve the equation
$$-\frac{T}{4\pi}\partial^2w_m
+f\frac{\partial{\cal F}}{\partial w_m}
=\frac{i}{2}\,\delta({\bf x}\!-\!{\bf z_1})\delta_{mm_1}$$
and plug the solution into the action \ur{action}. The solution is
$w_m({\bf x})\approx \delta_{mm_1}(i/2T)/|{\bf x}-{\bf z}_1|$ near the source where the
Laplacian is the leading term and $\partial{\cal F}/\partial w_m$ can be
neglected. At large distances from the source $w_m$ decays, therefore
$\partial{\cal F}/\partial w_m$ can be expanded to the linear order in $w_m$.
The solution decreases exponentially with the distance. At intermediate
distances the non-linearity is essential. However, whatever is the precise form
of the solution of this non-linear equation, the action is finite and independent
of $m_1$, as there is a perfect cyclic symmetry in $m_1$. Therefore, the action
factors out from the summation over $m_1$, and we obtain
\beq
<\!\Tr L\!>={\rm const.}\,\sum_{m_1}\exp\left(2\pi i \mu_{m_1}\right)=0,
\la{avL}\eeq
as expected in the confining phase. We use here the ``maximally-non-trivial''
holonomy \ur{muconf} which has been shown in Section IV to bring the free
energy to the minimum.

\subsection{Heavy quark potential}

The correlation function of two Polyakov lines in the
fundamental representation at spatial points ${\bf z}_1$ and ${\bf z}_2$ is
\bea\la{corr1}
&&\left<\Tr L({\bf z}_1)\Tr L^\dagger({\bf z}_2)\right>
=\sum_{m_1,n_1}e^{2\pi i(\mu_{m_1}-\mu_{n_1})}
\int\!Dw_m\exp\left(\!\int\!d{\bf x}\,\frac{4\pi f}{N}{\cal F}(w)\right)\\
\n
&&\cdot\!\prod_m\delta\left(-\frac{T}{4\pi}\partial^2w_m+f\frac{\partial{\cal F}}{\partial w_m}
-\frac{i}{2}\,\delta({\bf x}\!-\!{\bf z_1})\delta_{mm_1}
+\frac{i}{2}\,\delta({\bf x}\!-\!{\bf z_2})\delta_{mn_1}\right)\,
\det\left(-\frac{T}{4\pi}\partial^2\delta_{mn}+
f\frac{\partial^2{\cal F}}{\partial w_m\partial w_n}\right).
\eea

We are interested in the asymptotics of the correlator \ur{corr1}
at large source separations, $|{\bf z}_1\!-\!{\bf z}_2|\to\infty$.
We shall see in a moment that $w_m$'s solving the $\delta$-functions fall off
exponentially from the sources, $w\sim \exp(-M|{\bf z}\!-\!{\bf z}_{1,2}|)/|{\bf z}\!-\!{\bf z}_{1,2}|$,
therefore the generally non-linear equations on $w_m$ can be linearized far from the
sources. The same Yukawa (or, more precisely, Coulomb) functions are the solutions
close to the sources, as the leading term there is the Laplacian, and the
$\partial {\cal F}/\partial w_m$ term can be neglected. In the intermediate range
the non-linearity is essential but it has no influence on the asymptotics of
the potential between two infinitely heavy quarks, -- only on the residue
of the correlator. The action acquires the $|{\bf z}_1\!-\!{\bf z}_2|$-dependent
contribution from the range of integration far away from both sources where $w_m(x)$
is small. Therefore, to find the asymptotics of the heavy-quark potential one can take
$\partial{\cal F}/\partial w_m$ to the linear order in $w_m$ and ${\cal F}(w)$ to the
quadratic order. We have for small $w_m$
\beq
{\cal F}(w)=\sum_m e^{w_m-w_{m\!+\!1}}\approx N+\frac{1}{2}\, w_m\,{\cal M}_{mn}\,w_n,
\qquad \frac{\partial {\cal F}}{\partial w_m}\approx {\cal M}_{mn}\,w_n,
\la{calFsm}\eeq
where ${\cal M}$ is the matrix made of scalar products of the simple roots of
the gauge group, supplemented by a non-simple root to make it periodic: ${\cal M}_{mn}
=\Tr\,C_mC_n$, see \eq{EB}. In our case of $SU(N)$
\beq
{\cal M}=\left(\begin{array}{cccccc}2&-1&0&\ldots&0 & -1\\ -1&2&-1&\ldots &0& 0\\
0&-1&2&-1&\ldots &0\\ \ldots & \ldots & \ldots & \ldots & \ldots & \ldots\\-1&0&0&\ldots
&-1&2\end{array}\right).
\la{calM}\eeq
The $SU(2)$ group is a special case where this matrix is
\beq
{\cal M}^{(2)}=\left(\begin{array}{cc}2&-2\\-2&2\end{array}\right).
\la{calM2}\eeq
The ortho-normalized eigenvectors are the pairs
\beq
V^{(k,1)}_n=\sqrt{\frac{2}{N}}\cos\left(\frac{2\pi k}{N}\,n\right)\quad{\rm and}\quad
V^{(k,2)}_n=\sqrt{\frac{2}{N}}\sin\left(\frac{2\pi k}{N}\,n\right)
\la{eigv1}\eeq
corresponding to the twice degenerate eigenvalues
\beq
{\cal M}^{(k)}=\left(2\sin\frac{\pi k}{N}\right)^2,\quad k=1...\left[\frac{N\!-\!1}{2}\right].
\la{eig1}\eeq
There is also an eigenvector $V^{(0)}_n=\cos(2\pi\!\cdot\!0/N\!\cdot\!n)/\sqrt{N}=(1,1,...,1)/\sqrt{N}$
with a non-degenerate zero eigenvalue, and in case of even $N$ there is an additional
eigenvector $V^{(N/2)}_n=\cos(2\pi\!\cdot\!N/2/N\!\cdot\!n)/\sqrt{N}=(1,-1,1,...,-1)/\sqrt{N}$
with a non degenerate eigenvalue equal $2^2$. In other words, the eigenvalues are
\beq
{\cal M}^{(k)}=\left(2\sin\frac{\pi k}{N}\right)^2,\quad k=0,...,N-1,
\la{eig2}\eeq
where the pairs of eigenvalues corresponding to $k$ and $N-k$ are apparently
degenerate.

In the linearized form, the $\delta$-functions in \eq{corr1} impose the
equations
\beq
-\partial^2w_m+M^2{\cal M}_{mn}w_n=
\frac{2\pi i}{T}\left(\delta_{mm_1}\delta({\bf x}\!-\!{\bf z}_1)
-\delta_{mn_1}\delta({\bf x}\!-\!{\bf z}_2)\right)
\la{UDPollin}\eeq
where we have introduced the `dual photon' mass
\beq
M^2=\frac{4\pi f}{T}=\frac{16\pi^2\Lambda^4}{g^4 T^2}={\cal O}(N^2).
\la{M}\eeq
Equations \ur{UDPollin} are best solved in the momentum space:
\beq
w_m({\bf p})=\frac{2\pi i}{T}\left(\frac{1}{{\bf p}^2+M^2{\cal M}}\right)_{mn}\,
E_n({\bf p}),
\quad {\rm where}\quad
E_n=\delta_{nm_1}e^{i{\bf p}\cdot{\bf z}_1}
-\delta_{nn_1}e^{i{\bf p}\cdot{\bf z}_2}.
\la{wp}\eeq
This must be put into the action \ur{action} where ${\cal F}(w)$ is to be expanded
to the quadratic order. We have
\bea\n
&&\int\!d^3{\bf x}\,\frac{4\pi f}{N}\,\frac{1}{2}\,w_m({\bf x})
{\cal M}_{mn}w_n({\bf x})
=\frac{2\pi f}{N}\int\!\frac{d^3{\bf p}}{(2\pi)^3}\,w_m({\bf p}){\cal M}_{mn}w_n(-{\bf p})\\
\n
&&=-\frac{2\pi f}{N}\,\frac{(2\pi)^2}{T^2}
\int\!\frac{d^3{\bf p}}{(2\pi)^3}\,E_m({\bf p})
\left(\frac{1}{{\bf p}^2+M^2{\cal M}}\right)_{mp}\,{\cal M}_{pq}\,
\left(\frac{1}{{\bf p}^2+M^2{\cal M}}\right)_{qn}\,E_n(-{\bf p})\\
&&=-\frac{(2\pi)^3f}{NT^2}\int\!\frac{d^3{\bf p}}{(2\pi)^3}\,
E_m({\bf p})\sum_{l=1}^N V^{(l)}_m\,\frac{1}{{\bf p}^2+M^2{\cal M}^{(l)}}\,{\cal M}^{(l)}
\,\frac{1}{{\bf p}^2+M^2{\cal M}^{(l)}}\,V^{(l)}_n\,E_n(-{\bf p})
\la{corr2}\eea
where we have diagonalized the matrices by the orthogonal transformation built
of the eigenvectors $V^{(l)}_m$ corresponding to the eigenvalues ${\cal M}^{(l)}$.
We now pick from $E_m({\bf p})E_n(-{\bf p})$ the cross terms depending on ${\bf z}_1\!-\!{\bf z}_2$
as only they are relevant for the interaction. The inverse Fourier transform is
$$\int\!\frac{d^3{\bf p}}{(2\pi)^3}\,\frac{e^{i{\bf p}\cdot({\bf z}_1-{\bf z}_2)}}
{({\bf p}^2+M^2{\cal M}^{(l)})^2}=\frac{1}{8\pi}\,
\frac{e^{-|{\bf z}_1-{\bf z}_2|M\sqrt{{\cal M}^{(l)}}}}{M\sqrt{{\cal M}^{(l)}}}.$$
Therefore, we continue the chain of eqs.({\ref{corr2}}) and write
\beq
({\ref{corr2}})=\frac{2\pi^2 f}{NT^2 M}\,V^{(l)}_{m_1}\,\sqrt{{\cal M}^{(l)}}\,
e^{-|{\bf z}_1-{\bf z}_2|M\sqrt{{\cal M}^{(l)}}}\,V^{(l)}_{n_1}
\la{corr21}\eeq
where summation over all eigenvalues labeled by $l$ is implied.
The coefficient -2 arises because cross terms in $E_mE_n$ have negative relative sign,
and there are two such terms. We obtain from \eq{corr1}
\beq
\left<\Tr L({\bf z}_1)\Tr L^\dagger({\bf z}_2)\right>
=\sum_{m_1,n_1}\exp\left(2\pi i(\mu_{m_1}\!-\!\mu_{n_1})\!
+\!\frac{2\pi^2 f}{NT^2 M}\,V^{(l)}_{m_1}\,\sqrt{{\cal M}^{(l)}}\,
e^{-|{\bf z}_1-{\bf z}_2|M\sqrt{{\cal M}^{(l)}}}\,V^{(l)}_{n_1}\right)\,.
\la{corr3}\eeq

At large separations $|{\bf z}_1\!-\!{\bf z}_2|$ between the point sources
the second term in \eq{corr3} is exponentially small and one can Taylor-expand
it. The zero-order term is zero(!) as it is the product of two independent sums
over $m_1$ and $n_1$, {\it i.e.} it is the product of unconnected
$<\!\!\Tr L({\bf z}_1)\!\!><\!\!\Tr L^\dagger({\bf z}_2)\!\!>=0$, as explained in the
previous subsection. In the first non-zero order we get
\beq
\left<\Tr L({\bf z}_1)\Tr L^\dagger({\bf z}_2)\right>
=\frac{2\pi^2 f}{NT^2 M}\,\sum_{l=1}^{\left[\frac{N}{2}\right]}
\sqrt{{\cal M}^{(l)}}\,e^{-|{\bf z}_1-{\bf z}_2|M\sqrt{{\cal M}^{(l)}}}\,
\sum_{m_1,n_1=1}^N\exp\left(2\pi i(\mu_{m_1}\!-\!\mu_{n_1})\right)\,
V^{(l)}_{m_1}\,V^{(l)}_{n_1}\,.
\la{corr4}\eeq
It is a sum of exponentially decaying contributions with the exponents
determined by the eigenvalues ${\cal M}^{(l)}$, see \eq{eig2}. The weight
of the $l^{\rm th}$ contribution is determined by the summation over $m_1,n_1$.
For $l=1,...,\left[\frac{N\!-\!1}{2}\right]$, eigenvalues are twice degenerate
and we use the eigenvectors \ur{eigv1}. At even $N$ the highest eigenvalue is
non-degenerate, the corresponding eigenvector being $V^{(N/2)}_n=\cos \pi n/\sqrt{N}$.
Summation over $m_1,n_1$ in \eq{corr4} gives
\bea\la{ortho1}
\sum_{m_1,n_1=1}^N\!\!\!\!&&\!\!\!\!\!\exp\!\left(\!2\pi i\frac{m_1\!-\!n_1}{N}\!\right)\,
\frac{2}{N}\left(\!\cos\frac{2\pi l m_1}{N}\cos\frac{2\pi l n_1}{N}
\!+\!\sin\frac{2\pi l m_1}{N}\sin\frac{2\pi l n_1}{N}\right)\\
\n
&=&\!N\,\delta_{l,1}\,,\quad
{\rm for\;all}\;l\!=\!1...\left[\frac{N\!-\!1}{2}\right]\!,\;{\rm any}\;N;\\
\n
\sum_{m_1,n_1=1}^N\!\!\!\!&&\!\!\!\!\!\exp\!\left(\!2\pi i\frac{m_1\!-\!n_1}{N}\!\right)\,
\frac{1}{N}\,\cos\pi m_1\,\cos\pi n_1\\
\n
&=&N\,\delta_{N,2}\,,\quad {\rm for}\;l=\frac{N}{2},\;{\rm even}\;N.
\eea
We see that only the exponent with the {\em lowest} eigenvalue
$\surd{\cal M}^{(1)}=2\,\sin\frac{\pi}{N}$ contributes in \eq{corr3}
to the correlator of Polyakov lines in the fundamental representation;
higher eigenvalues decouple through orthogonality. We thus obtain
\beq
\left<\Tr L({\bf z}_1)\Tr L^\dagger({\bf z}_2)\right>
=\frac{2\pi^2 f}{NT^2 M}\,2\sin\frac{\pi}{N}\,
N\,\exp\left(-|{\bf z}_1\!-\!{\bf z}_2|\,M\,2\sin\frac{\pi}{N}\right),
\la{corr5}\eeq
plus exponentially small corrections from the expansion of \eq{corr3}
to higher orders. This should be compared with the standard definition of the
heavy-quark potential
$$
\left<\Tr L({\bf z}_1)\Tr L^\dagger({\bf z}_2)\right>
=C\,\exp\left(-\frac{V({\bf z}_1\!-\!{\bf z}_2)}{T}\right)
$$
from where we deduce the linear heavy-quark potential at large separations:
\beq
V({\bf z}_1\!-\!{\bf z}_2)=|{\bf z}_1\!-\!{\bf z}_2|\,MT\,2\sin\frac{\pi}{N}
=\sigma\,|{\bf z}_1\!-\!{\bf z}_2|,\qquad C={\cal O}(N^0),
\la{V}\eeq
with the `string tension'
\beq
\sigma=MT\,2\sin\frac{\pi}{N}=T\sqrt{\frac{4\pi f}{T}}\,2\sin\frac{\pi}{N}
=8\pi\,\frac{\Lambda^2}{g^2}\,\sin\frac{\pi}{N}
=\frac{\Lambda^2}{\lambda}\,\frac{N}{\pi}\,\sin\frac{\pi}{N}\,.
\la{sigmaE}\eeq
In the last equation in the chain the $N$-independent 't Hooft coupling $\lambda$ has
been used. We see that the string tension turns out to be {\it i}) independent of the
temperature~\cite{footnote2} and {\it ii}) independent of $N$ at large $N$, as expected.
In reality we expect that anti-self-dual dyons not accounted for here double $M^2$ and
hence the dyon-induced string tension is actually $\sqrt{2}$ times bigger.
A more robust quantity (both from the theoretical and lattice viewpoints) is
the ratio $T_c\sqrt{\sigma}$ since in this ratio the poorly known parameters
$\Lambda$ and $\lambda$ cancel out, see \eq{Tcrit}:
\beq
\frac{T_c}{\sqrt{\sigma}}=\left(\frac{45}{4\pi^4}\,\frac{\pi^2 N^2}{(N^4-1)\sin^2\frac{\pi}{N}}
\right)^{\frac{1}{4}}\quad\stackrel{N\to\infty}{\longrightarrow}\quad
\frac{1}{\pi}\left(\frac{45}{4}\right)^{\frac{1}{4}}+{\cal O}\left(\frac{1}{N^2}\right).
\la{ratio}\eeq

The values are compared to those measured in lattice simulations of the pure $SU(N)$ gauge
theories~\cite{Teper} in Table~1 demonstrating a good agreement. The relatively large
4\% deviation for the $SU(2)$ group may be related to the fact that we have determined
$T_c$ in Section IV by comparing the free energy for confining and trivial holonomy,
that is assuming a first-order transition, whereas for $N\!=\!2$ it is actually
a second order one.

\begin{table}[h]
\begin{tabular}{|c|c|c|c|c|c|}
\hline
$N\!=\!2$ & 3 & 4 & 6 & 8 & $\infty$ \\
\hline
&&&&&\\
 (0.7425) & 0.6430 & 0.6150 & 0.5967 & 0.5906 &
$0.5830+\frac{0.4795}{N^2}+\frac{0.5006}{N^4}+...$\\
&&&&&\\
\hline
&&&&&\\
 0.7091(36) & 0.6462(30) & 0.6344(81) & 0.6101(51) & 0.5928(107) &
$0.5970(38)+
\frac{0.449(29)}{N^2}\;({\rm fit})$\\
&&&&&\\
\hline
\end{tabular}
\caption{Deconfinement temperature $T_c/\sqrt{\sigma}$ from \eq{ratio} (upper row) and from lattice
simulations~\cite{Teper} (lower row).}
\end{table}

In Table II we add the comparison of the topological susceptibility \ur{topsusc} measured in units
of the string tension, with the lattice data~\cite{TeperQT}. The agreement is also remarkably good,
given the approximate nature of the model.

\begin{table}[h]
\begin{tabular}{|c|c|c|c|c|}
\hline
$N\!=\!2$ & 3 & 4 & 5 & $\infty$ \\
\hline
&&&&\\
 0.5 & 0.439 & 0.420 & 0.412 &
$0.399+\frac{0.328}{N^2}+\frac{0.243}{N^4}+...$\\
&&&&\\
\hline
&&&&\\
 0.4831(56) & 0.434(10) & 0.387(17) & 0.387(21) &
$0.376(20)+
\frac{0.43(10)}{N^2}\;({\rm fit})$\\
&&&&\\
\hline
\end{tabular}
\caption{Topological susceptibility $<\!Q_T^2\!>^{1/4}/\sqrt{\sigma}$ from \eq{topsusc}
(upper row) and from lattice simulations~\cite{TeperQT} (lower row).}
\end{table}

\newpage

\subsection{$N$-ality and $k$-strings}

All irreducible representations of the $SU(N)$ group fall into $N$ classes:
those that appear in the direct product of any number of adjoint representations,
and those that appear in the direct product of any number of adjoint representations
with the irreducible representation being the rank-$k$ antisymmetric tensor,
$k=1,\ldots , N\!-\!1$. ``$N$-ality'' is said to be zero in the first case and
equal to $k$ in the second. $N$-ality-zero representations transform trivially
under the center of the group $Z_N$; the rest acquire a phase $2\pi k/N$.

One expects that there is no asymptotic linear potential between static color
sources in the adjoint representation as such sources are screened by gluons.
If a representation is found in a direct product of some number of adjoint
representations and a rank-$k$ antisymmetric representation, the adjoint ones
``cancel out'' as they can be all screened by an appropriate number of gluons.
Therefore, from the confinement viewpoint all $N$-ality $=k$ representations are
equivalent and there are only $N-1$ string tensions $\sigma(k,N)$ being the
coefficients in the asymptotic linear potential for sources in the antisymmetric
rank-$k$ representation. Its dimension is $d(k,N)=\frac{N!}{k!(N-k)!}$ and the
eigenvalue of the quadratic Casimir operator is $C(k,N)=\frac{N+1}{2N}\,k(N-k)$.

The value $k\!=\!1$ corresponds to the fundamental representation whereas
$k=N\!-\!1$ corresponds to the representation conjugate to the fundamental
[quarks and anti-quarks]. In general, the rank-$(N\!-\!k)$ antisymmetric
representation is conjugate to the rank-$k$ one; it has the same dimension
and the same string tension, $\sigma(k,N)=\sigma(N\!-\!k,N)$. Therefore, for
odd $N$ all string tensions appear in equal pairs; for even $N$, apart from
pairs, there is one privileged representation with $k=\frac{N}{2}$ which
has no pair and is real. The total number of different string tensions
is thus $\left[\frac{N}{2}\right]$.

The behavior of $\sigma(k,N)$ as function of $k$ and $N$ is an important issue as
it discriminates between various confinement mechanisms. On general $N$-counting
grounds one can only infer that at large $N$ and $k\ll N$, $\sigma(k,N)/\sigma(1,N)=
(k/N)(1+{\cal O}(1/N^2))$~\cite{Shifman}. In this subsection
we show that the dyon ensemble leads to the sine law for the $k$-strings,
\beq
\sigma(k,N)={\rm const.}\,\sin\frac{\pi k}{N}
\qquad({\rm plus\;temperature\;dependent\;corrections})
\la{sine}\eeq
satisfying the above requirement on the asymptotics. The sine behavior has been
found in certain supersymmetric theories~\cite{sine}. Here it follows from a
direct calculation of the correlator of Polyakov lines in the rank-$k$
antisymmetric representation.

We first show that there is no asymptotic linear potential between adjoint
sources. If $A_4$ is diagonal and given by \eq{Av} the eigenvalues of the
Polyakov loop in the adjoint representation are
$\exp\left(\pm i(A_{4m}\!-\!A_{4n})/T\right)$, and there are $N\!-\!1$ unity eigenvalues.
Therefore, the average of the adjoint line is non-zero, and the correlator of
two such lines tends asymptotically to a non-zero constant.

Let the Polyakov line in the fundamental representation be $L({\bf z})
=\exp(iA_4({\bf z})/T)={\rm diag}(z_1,z_2,\ldots,z_N)$ where
$z_m=\exp\left(2\pi i\mu_m-\frac{i}{2}v_m({\bf z})\right)$, see \eq{Av}.
The Polyakov lines in the antisymmetric rank-$k$ representation are then
\bea\n
L(1,N)&=&\Tr L = \sum_{m=1}^Nz_m,\qquad k=1,\\
\n
L(2,N)&=&\frac{1}{2}\left((\Tr L)^2-\Tr L^2\right) = \sum_{m<n}^Nz_mz_n,\qquad k=2,\\
\n
L(3,N)&=&\frac{1}{6}\left((\Tr L)^3-3\Tr L^2\,\Tr L+2 \Tr L^3\right)
= \sum_{m<n<p}^Nz_mz_nz_p,\qquad k=3,\\
L(k,N)&=&\sum_{m_1<m_2<...<m_k}^Nz_{m_1}z_{m_2}...z_{m_k}.
\la{Lk}\eea
Therefore, any general $L(k,N)({\bf z})$ placed at the $3d$ point ${\bf z}$
serves as a source
$$
\sum_{m_1<m_2<...<m_k}^N
\exp\left[2\pi i\left(\mu_{m_1}\!+...+\!\mu_{m_k}\right)\!-\frac{i}{2}
\left(v_{m_1}({\bf z})\!+...+\!v_{m_k}({\bf z})\right)\right]
$$
for the $v_m$ field, which should be put into the partition function \ur{Z5}.

To get the correlation function of two lines in $k$-representation we proceed
as in Subsection V.~B and arrive at the generalization of \eq{corr4}:
\bea\la{corrLk1}
&&\left<\Tr L(k,N)({\bf z}_1)\;\;\Tr L^\dagger(k,N)({\bf z}_2)\right>
=\frac{2\pi^2 f}{NT^2 M}\,\sum_{l=1}^{\left[\frac{N}{2}\right]}
\sqrt{{\cal M}^{(l)}}\,e^{-|{\bf z}_1-{\bf z}_2|M\sqrt{{\cal M}^{(l)}}}\\
\n
&&\!\!\cdot\!\!\sum_{m_1<m_2<...<m_k}^N\;\;\sum_{n_1<n_2<...<n_k}^N
\!\!\exp 2\pi i\left(\!\mu_{m_1}\!+...+\!\mu_{m_k}\!-\!\mu_{n_1}\!-...-\!\mu_{n_k}\!\right)
\left[V^{(l)}_{m_1}+...+V^{(l)}_{m_k}\right]\,
\left[V^{(l)}_{n_1}+...+V^{(l)}_{n_k}\right]\,.
\eea
In deriving \eq{corrLk1} it is important that the maximally non-trivial
holonomy \ur{muconf} is used, leading to $<\!\Tr L(k,N)\!>=0,\;k=1...N\!-\!1$.
Higher powers of $\exp\left(-|{\bf z}_1\!-\!{\bf z}_2|M\sqrt{{\cal M}^{(l)}}\right)$
have been neglected.

Again, the correlation function of Polyakov lines is a sum of exponentially
decaying contributions with the exponents determined by the eigenvalues
${\cal M}^{(l)}$, see \eq{eig2}. The weight of the $l^{\rm th}$ contribution
is given by the sum over $m_{1,2...k}$ and $n_{1,2...k}$. We recall the eigenvectors
$V^{(l)}$ \ur{eigv1} and observe the following important orthogonality relation:
\bea\n
&&\sum_{m_1<m_2<...<m_k}^N\;\;\sum_{n_1<n_2<...<n_k}^N
\!\!\exp\left[2\pi i\frac{m_1\!+...+\!m_k\!-\!n_1\!-...-\!n_k}{N}\!\right]\,
\left[V^{(l)}_{m_1}+...+V^{(l)}_{m_k}\right]\,
\left[V^{(l)}_{n_1}+...+V^{(l)}_{n_k}\right]\\
\n\\
&&=\left\{\begin{array}{ccc}
N\,\delta_{lk} & {\rm for\;all\;twice\;degenerate\;eigenvalues}\;l=1\ldots\left[\frac{N-1}{2}\right], &
{\rm any}\;N\geq 2k,\\
&&\\
N\,\delta_{N,2k} & {\rm for\;the\;highest,\;non}\!-\!{\rm degenerate\;eigenvalue}\;
l=\frac{N}{2}, & {\rm even}\;N.\end{array}\right.
\la{ortho}\eea
[The orthogonality relation \ur{ortho1} is a particular case of this general
one, corresponding to $k\!=\!1$. The derivation of these relations
is elementary when one presents the eigenvectors in the exponential form.]

The above orthogonality relations imply that the correlator of the lines in rank-$k$
antisymmetric tensor representation {\em couples only to the single exponent} determined by the
$k^{\rm th}$ eigenvalue $\surd{\cal M}^{(k)}=2\sin\frac{\pi k}{N},\;N\geq 2k$;
all the rest eigenvalues decouple~\cite{footnote3}.
Therefore, the correlation function \ur {corrLk1} is
\beq
\left<\Tr L(k,N)({\bf z}_1)\;\;\Tr L^\dagger(k,N)({\bf z}_2)\right>
=\frac{2\pi^2 f}{NT^2 M}\,2\sin\frac{\pi k}{N}\,N\,
\exp\left(-|{\bf z}_1-{\bf z}_2|\,M\sqrt{{\cal M}^{(k)}}\right)
\la{corrLk2}\eeq
and hence the general-$k$ string tension is
\beq
\sigma(k,N)=MT\sqrt{{\cal M}^{(k)}}=MT\,2\sin\frac{\pi k}{N}
=\frac{\Lambda^2}{\lambda}\,\frac{N}{\pi}\,\sin\frac{\pi k}{N}
\la{sigma-k}\eeq
as announced. Lattice simulations~\cite{DelDebbio-k} support this regime,
whereas another lattice study~\cite{Teper-k} gives somewhat smaller values
but within two standard deviations from the values following from \eq{sigma-k}.
For a general discussion of the sine regime for $k$-strings, which is favored
from many viewpoints, see~\cite{Shifman}.

\section{Area law for spatial Wilson loops}

The area behavior of the {\em spatial} Wilson loops is not directly related to
the linear confining potential, however it is believed that in a confining
theory the spatial Wilson loop must exhibit the area law. The reason is that
{\it i}) at $T\to 0$ Lorentz symmetry is restored, therefore the spatial loop
must behave in the same way as the time-like one whose area law is related to
the linear confining potential, {\it ii}) at high $T$ the spatial loop eventually
becomes a time-like loop from the $2\!+\!1$ dimensions' point of view, which
has to obey the area law to fulfill confinement in $3d$. Therefore, it is very
plausible that the spatial Wilson loop has the area behavior at any temperatures.
It is expected that the spatial string tension is roughly constant below the
deconfinement transition, and eventually grows as $\sim\!T^2$ at very high
temperatures where the theory is basically 3-dimensional.

In this section we demonstrate that the dyon ensemble induces the area law
for spatial Wilson loops and that the string tension coincides with that
found in the previous section from the correlators of the Polyakov lines. We
think that it is an interesting result since a) the way we derive the string
tension for spatial loops is very different from that for Polyakov lines,
b) in a sense it demonstrates that our ensemble restores Lorentz symmetry at
low temperatures, despite its 3-dimensional formulation.

The condition that $A_4$ is time-independent only partially fixes the gauge:
one can still perform time-independent gauge transformations. This freedom
can be used to make $A_4$ diagonal ({\it i.e.} Abelian). This necessarily
implies Dirac string singularities which are pure gauge artifacts as they do
not carry any energy. Moreover, the Dirac strings' directions are also subject
to the freedom of the gauge choice. In Refs.~\cite{KvB,KvBSUN} the gauge choice
in the explicit construction of the KvBLL instanton was such that the Dirac
strings were connecting individual dyon constituents of the instanton. This
choice is, however, not convenient in the ensemble of dyons as dyons have to loose
their ``memory'' to what particular instanton they belong to. The natural gauge
is where all Dirac strings of all dyons are directed to infinity along some
axis, {\it e.g.} along the $z$ axis. The dyons' field in this gauge is given
explicitly in Ref.~\cite{DP-SUSY} (for the $SU(2)$ group).

In this gauge, the magnetic field of dyons beyond their cores is also Abelian
and is a superposition of the Abelian fields of individual dyons. For large
Wilson loops we are interested in, the field of a large number of dyons
contribute as they have a slowly decreasing $1/|{\bf x}\!-\!{\bf x}_i|$
asymptotics, hence the use of the field outside the cores is justified.
Owing to self-duality,
\beq
\left[B_i({\bf x})\right]_{mn}=\left[\partial_iA_4({\bf x})\right]_{mn}
=-\frac{T}{2}\,\delta_{mn}\,\partial_iv_m({\bf x}),
\la{B}\eeq
cf. \eq{vm}. Since $A_i$ is Abelian beyond the cores, one can use the Stokes
theorem for the spatial Wilson loop:
\beq
W\equiv\Tr\,{\rm P}\exp\,i\oint\!A_idx^i=\Tr\exp\,i\int\!B_i\,d^2\sigma^i
=\sum_m\exp\left(-i\frac{T}{2}\int\!d^2\sigma^i\,\partial_iv_m\right).
\la{Wi1}\eeq
\Eq{Wi1} may look contradictory as we first use $B_i={\rm curl}A_i$ and then
$B_i=\partial_iA_4$. Actually there is no contradiction as the last equation
is true up to Dirac string singularities which carry away the magnetic flux.
If the Dirac string pierces the surface spanning the loop it gives a quantized
contribution $\exp(2\pi i\!\cdot\!{\rm integer})=1$; one can also use the gauge
freedom to direct Dirac strings parallel to the loop surface in which case
there is no contribution from the Dirac strings at all.

Let us take a flat Wilson loop lying in the $(xy)$ plane at $z\!=\!0$. Then
\eq{Wi1} is continued as
\beq
W=\sum_m\exp\left(-i\frac{T}{2}\int_{x,y\in {\rm Area}}\!d^3x\,\partial_zv_m\delta(z)\right)
=\sum_m\exp\left(i\frac{T}{2}\int_{x,y\in {\rm Area}}\!d^3x\,v_m\,\partial_z\delta(z)\right)\,.
\la{Wi2}\eeq
It means that the average of the Wilson loop in the dyons ensemble is given by
the partition function \ur{Z5} with the source
$$
\sum_m\exp\left(i\frac{T}{2}\int\!d^3x\,v_m\,\frac{d\delta(z)}{dz}\,
\theta(x,y\in {\rm Area})\right)
$$
where $\theta(x,y\in {\rm Area})$ is a step function equal to unity if $x,y$ belong
to the area inside the loop and equal to zero otherwise. Again, the source shifts the
argument of the $\delta$-function arising from the integration over the $v_m$
variables, and the average Wilson loop in the fundamental representation is given
by the equation
\bea\la{Wi3}
\left<W\right>&=&\sum_{m_1}\int\!Dw_m\exp\left(\!\int\!d{\bf x}\,\frac{4\pi f}{N}{\cal F}(w)\right)\\
\n
&\cdot\!&\!\!\!\!\!\prod_m
\delta\left(\!-\frac{T}{4\pi}\partial^2w_m+f\frac{\partial{\cal F}}{\partial w_m}
+\frac{i T}{2}\,\delta_{mm_1}\,\frac{d\delta(z)}{dz}\,\theta(x,y\in {\rm Area})
\right)\,
\det\left(\!-\frac{T}{4\pi}\partial^2\delta_{mn}+
f\frac{\partial^2{\cal F}}{\partial w_m\partial w_n}\right).
\eea
Therefore, one has to solve the non-linear equations on $w_m$'s with
a source along the surface of the loop,
\beq
-\partial^2w_m+M^2\left(e^{w_m-w_{m+1}}-e^{w_{m-1}-w_m}\right)
=-2\pi i\,\delta_{mm_1}\,\frac{d\delta(z)}{dz}\,\theta(x,y\in {\rm Area}),
\qquad M^2=\frac{4\pi f}{T},
\la{UD3}\eeq
for all $m_1$, plug it into the action $(4\pi f/N){\cal F}(w)$, and sum over
$m_ 1$. In order to evaluate the average of the Wilson loop in a general antisymmetric
rank-$k$ representation, one has to take the source in \eq{UD3} as $-2\pi i\,\delta'(z)\,
\left(\delta_{mm_1}\!+\!\ldots\!+\!\delta_{mm_k}\right)$ and sum over $m_1\!<\!\ldots\!<\!m_k$
from 1 to $N$, see \eq{Lk}. Again, the ghost determinant cancels exactly the Jacobian from
the fluctuations of $w_m$ about the solution, therefore the classical-field calculation
is exact.

Contrary to the case of the Polyakov lines, one cannot, generally speaking,
linearize \eq{UD3} in $w_m$ but has to solve the non-linear equations as they are.
With no source in the r.h.s., \eq{UD3} is known as the periodic Toda lattice
and it is integrable for any $N$. It has an hierarchy of soliton solutions
constructed in Ref.~\cite{Toda,Toda-sol}. Below we modify those solutions in such a way
that they satisfy \eq{UD3} with a source in the r.h.s. We call them ``pinned solitons'';
their action determine the string tensions. We obtain below for the `magnetic' $k$-string tension
\beq
\sigma(k,N)=MT\,2\sin\frac{\pi k}{N}
=\frac{\Lambda^2}{\lambda}\,\frac{N}{\pi}\,\sin\frac{\pi k}{N},
\la{sigmaM-k}\eeq
which coincides exactly with the `electric' string tension \ur{sigma-k} found
from the correlators of the Polyakov lines.

\subsection{Construction of string solitons}

Let us find the pinned solitons corresponding to Wilson loops in a general
antisymmetric rank-$k$ representation of the $SU(N)$ gauge group.
First of all we rewrite \eq{UD3} for the difference fields $w_{m,m+1}=w_m-w_{m+1}$ as only
the differences enter the action:
\bea\la{UD4}
&&-w''_{12}+M^2\left(2e^{w_{12}}-e^{w_{N1}}-e^{w_{23}}\right)
=-2\pi i\delta'(z)\,\left[\left(\delta_{1,m_1}\!+\!\ldots\!+\!\delta_{1,m_k}\right)
-\left(\delta_{2,m_1}\!+\!\ldots\!+\!\delta_{2,m_k}\right)\right],\\
\n
&&-w''_{23}+M^2\left(2e^{w_{23}}-e^{w_{12}}-e^{w_{34}}\right)
=-2\pi i\delta'(z)\,\left[\left(\delta_{2,m_1}\!+\!\ldots\!+\!\delta_{2,m_k}\right)
-\left(\delta_{3,m_1}\!+\!\ldots\!+\!\delta_{3,m_k}\right)\right],\\
\n
&&\ldots\\
\n
&&-w''_{N-1,N}+M^2\left(2e^{w_{N-1,N}}-e^{w_{N-2,N-1}}-e^{w_{N,1}}\right)
=-2\pi i\delta'(z)\,\left[\left(\delta_{N-1,m_1}\!+\!\ldots\!+\!\delta_{N-1,m_k}\right)
-\left(\delta_{N,m_1}\!+\!\ldots\!+\!\delta_{N,m_k}\right)\right],\\
\n
&&-w''_{N,1}+M^2\left(2e^{w_{N,1}}-e^{w_{N-1,N}}-e^{w_{12}}\right)
=-2\pi i\delta'(z)\,\left[\left(\delta_{N,m_1}\!+\!\ldots\!+\!\delta_{N,m_k}\right)
-\left(\delta_{1,m_1}\!+\!\ldots\!+\!\delta_{1,m_k}\right)\right]\,.
\eea
At all $z$ except one point $z\!=\!0$ where there is a source, $w_{m,m+1}$ satisfy
free (zero source) equations. Solutions for very similar equations have been constructed in
Ref.~\cite{Toda,Toda-sol}. Adjusting them to our case, we write the general solutions of
\eq{UD4} with the zero r.h.s.:
\bea\la{genw}
&&w^{(k)}_{m,m+1}(z)=\ln\,\frac{\left[1+\gamma \varkappa^{k(m-1)}E^{(k)}(z)\right]
\left[1+\gamma \varkappa^{k(m+1)}E^{(k)}(z)\right]}{\left[1+\gamma \varkappa^{k\,m}E^{(k)}(z)\right]^2}\,,
\qquad k=1,\ldots ,N\!-\!1,\\
\n
&&E^{(k)}(z)=\exp(-M\sqrt{{\cal M}^{(k)}}\,z),\quad\sqrt{{\cal M}^{(k)}}=2\sin\frac{\pi k}{N},
\quad \varkappa=\exp\left(\pm\frac{2\pi i}{N}\right)\,.
\eea
The complex parameter $\gamma$ is arbitrary; there is also freedom in choosing the sign of the phase
of $\varkappa$. Equations \ur{genw} describe free ``domain wall'' solitons on the axis $-\infty <z<+\infty$;
it is easy to check that their actions do not depend on $\gamma$ and are given by \eq{sigmaM-k}.
Were $\gamma$ real it could be understood as an overall shift of the domain wall by $z_0$ where
$E^{(k)}(z_0)=\gamma$. However, we need pinned soliton solutions with a non-zero $\delta'(z)$
source in the r.h.s. To find these, we have to take one free solution at $z>0$ and another free solution at $z<0$,
where the two solutions may differ only by the value of the so far arbitrary complex parameter $\gamma$,
and by the sign of the phase of $\varkappa$.

We fix $\varkappa=\exp\left(+\frac{2\pi i}{N}\right)$ at $z>0$, and $\varkappa=\exp\left(-\frac{2\pi i}{N}\right)$
at $z<0$. The condition that only the imaginary parts of $w_{m,m+1}$ have jumps
at $z=0$ requires that $\gamma$ is a pure phase factor, $\gamma=e^{i\alpha}$
at $z>0$, and $\gamma=e^{-i\alpha}$ at $z<0$. Indeed, with these definitions, $w^{(k)}_{m,m+1}$
at $z<0$ are complex conjugates of the same functions at $z>0$; the real parts of $w^{(k)}_{m,m+1}$ are
continuous functions at $z=0$ whereas the imaginary parts may have jumps. We define the
logarithms in \eq{genw} such that they have cuts along the negative axis. Thus, the general form
of a ``pinned'' soliton solving \eq{UD4} with a non-zero source is
\beq
w^{(k)}_{m,m+1}(z)=\left\{\begin{array}{cc}
\ln\,\frac{\left[1+\gamma \varkappa^{k(m-1)}E^{(k)}(z)\right]
\left[1+\gamma \varkappa^{k(m+1)}E^{(k)}(z)\right]}{\left[1+\gamma \varkappa^{k\,m}E^{(k)}(z)\right]^2}\,,
& z>0,\\
&\\
\ln\,\frac{\left[1+\gamma^*\varkappa^{*\,k(m-1)}E^{(k)}(-z)\right]
\left[1+\gamma^*\varkappa^{*\,k(m+1)}E^{(k)}(-z)\right]}
{\left[1+\gamma^*\varkappa^{*\,k\,m}E^{(k)}(-z)\right]^2}\,,
& z<0, \end{array}\right.\qquad m=1,\ldots ,N,
\qquad k=1,\ldots ,N\!-\!1.
\la{pinned_soliton}\eeq
We have now to choose the phase factor $\gamma$ such that the functions
$w^{(k)}_{m,m+1}(z)$ have $\pm 2\pi i$ jumps at $z\!=\!0$ in accordance with
the source in the r.h.s. of \eq{UD4}. We note that at $z\to\pm\infty$ the arguments of
the logarithms tend to $e^{i\cdot 0}$, hence all functions tend to zero at $z\to\pm\infty$.
As one varies $|z|$ from $\infty$ to $0$, the arguments of the logarithms travel in the
complex plane, ending up at the {\it real axis} at $z\to 0$. The trajectories for $z>0$ and for $z<0$
are mirror images of one another since $w^{(k)}_{m,m+1}(-z)=\left(w^{(k)}_{m,m+1}(z)\right)^*$.
If at $z\to 0$ the trajectories end up at the positive semi-axis, the function has no
jump since the logarithm is uniquely defined there. If at $z\to 0$ the trajectories end up
at the negative semi-axis, the function has a $\pm 2\pi i$ jump owing to the cut of the logarithm
along the negative semi-axis. The sign of the jump depends on whether the trajectory approaches
the cut from above or from below. For given $N$ and $k$, the only handle ruling the behavior
of the trajectories in the complex plane is the phase factor $\gamma$. We shall show below that
one can find $\gamma$ such that a given function $w_{m,m+1}$ has a needed jump. But before
presenting explicit solutions for $k=1,2,\ldots$ let us show that the string tension for
a general $k$ representation is given by \eq{sigmaM-k}.

To find the string tension, one needs to compute the action on the solution \ur{pinned_soliton}:
$$
{\rm Action}(k,N)=\frac{4\pi f}{N}\int\!d^3x\sum_{m=1}^N\left[\exp\left(w^{(k)}_{m,m+1}\right)-1\right]
\,\theta(x,y\in {\rm Area})=-\sigma(k,N)\,{\rm Area}
$$
where we have subtracted the constant related to the vacuum. It is understood that the
solution \ur{pinned_soliton} is valid for $x,y$ inside the loop because
of the $\theta$-function in \ur{UD3} that we have omitted for brevity; outside the
loop there is no source and $w_{m,m+1}(z)=0$ is compatible with the equation. At the
loop boundary $w_{m,m+1}(z)$ interpolates between \ur{pinned_soliton} and zero.
Substituting the solution \ur{pinned_soliton} we obtain for the string tension
$$
\sigma(k,N)=-\frac{4\pi f}{N}
\int_{-\infty}^{\infty}dz\,\sum_{m=1}^N\left[\exp\left(w^{(k)}_{m,m+1}(z)\right)\!-\!1\right]
=\frac{4\pi f}{N}\sum_{m=1}^N\int_{-\infty}^{\infty}dz\,
\left(\varkappa^k\!-\!2\!+\!\varkappa^{-k}\right)\frac{\gamma\varkappa^{km} E^{(k)}(z)}
{\left(1\!+\!\gamma\varkappa^{km} E^{(k)}(z)\right)^2}
$$
where for $z<0$ one has to change $E^{(k)}(z)\to E^{(k)}(-z)=1/E^{(k)}(z),\,
\varkappa\to \varkappa^*=1/\varkappa,\,\gamma\to\gamma^*=1/\gamma$.
In fact the integrand is invariant under such change,
therefore one can proceed with the above expression integrating from $-\infty$ to $+\infty$:
the integral is equal $1/(M\surd{\cal M}^{(k)})$ and does not depend either on
$m$ or $\gamma$. Therefore there are $N$ equal terms in the sum and we
obtain finally the string tension
$$
\sigma(k,N)=4\pi f\,\frac{\varkappa^k-2+\varkappa^{-k}}{2M\sin\frac{\pi k}{N}}
=MT\,2\,\sin\frac{\pi k}{N}
$$
as announced.

\subsection{Wilson loop in the fundamental representation, $k=1$}

It is easy to verify that if one takes ${\rm arg}(\gamma)$ inside one of the $N$ equal-length
intervals covering the whole $2\pi$ range,
\beq
{\rm arg}(\gamma)\in\quad\left(\pi,\,\pi-\frac{2\pi}{N}\right),\quad
\left(\pi-\frac{2\pi}{N},\,\pi-\frac{4\pi}{N}\right),\quad
\left(\pi-\frac{4\pi}{N},\,\pi-\frac{6\pi}{N}\right),\ldots\,,
\left(-\pi+\frac{2\pi}{N},\,-\pi\right)\,,
\la{argg1int}\eeq
\eq{pinned_soliton} for $k\!=\!1$ gives the solutions of \eq{UD4} corresponding
to a single source at $m_1\!=\!1,2,3,\ldots,N$, respectively. For example, taking ${\rm arg}(\gamma)$
inside the first interval makes the functions $w_{12}$ and $w_{N,1}$ discontinuous at $z=0$
where their imaginary parts have a $2\pi$ jump in accordance with the source term
$2\pi i\,\delta'(z)\,\delta_{m1}$. All the rest functions are continuous. When one moves
${\rm arg}(\gamma)$ to the second interval in \ur{argg1int}, the functions $w_{12}$ and
$w_{23}$ have jumps in accordance with the source term $2\pi i\,\delta'(z)\,\delta_{m2}$
while all other functions are continuous, and so on. An example of the solutions for
$N\!=\!3$ is shown in Fig.~1 where ${\rm arg}(\gamma)$ is taken from the first interval,
in this case $\left(\pi,\frac{\pi}{3}\right)$. When ${\rm arg}(\gamma)$ is taken from
the second interval $\left(\frac{\pi}{3},-\frac{\pi}{3}\right)$ or from the third interval
$\left(-\frac{\pi}{3},-\pi,\right)$ the functions change cyclically $w_{12}\to
w_{23}\to w_{31}\to w_{12}$.

\begin{figure}[htb]
\includegraphics[width=0.28\textwidth]{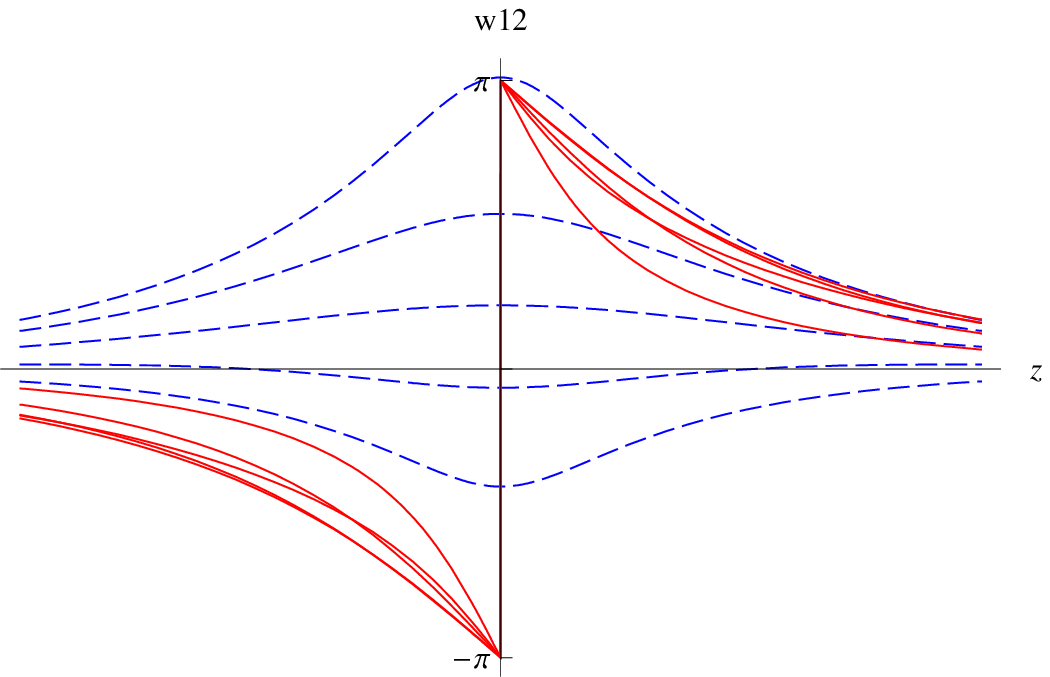}
\includegraphics[width=0.28\textwidth]{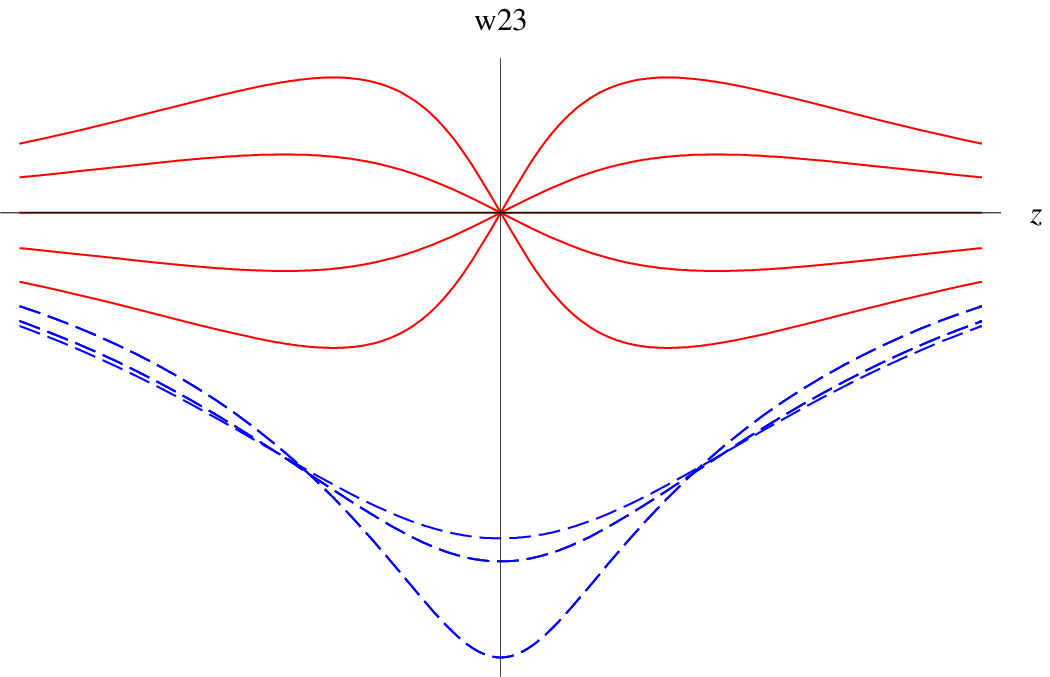}
\includegraphics[width=0.28\textwidth]{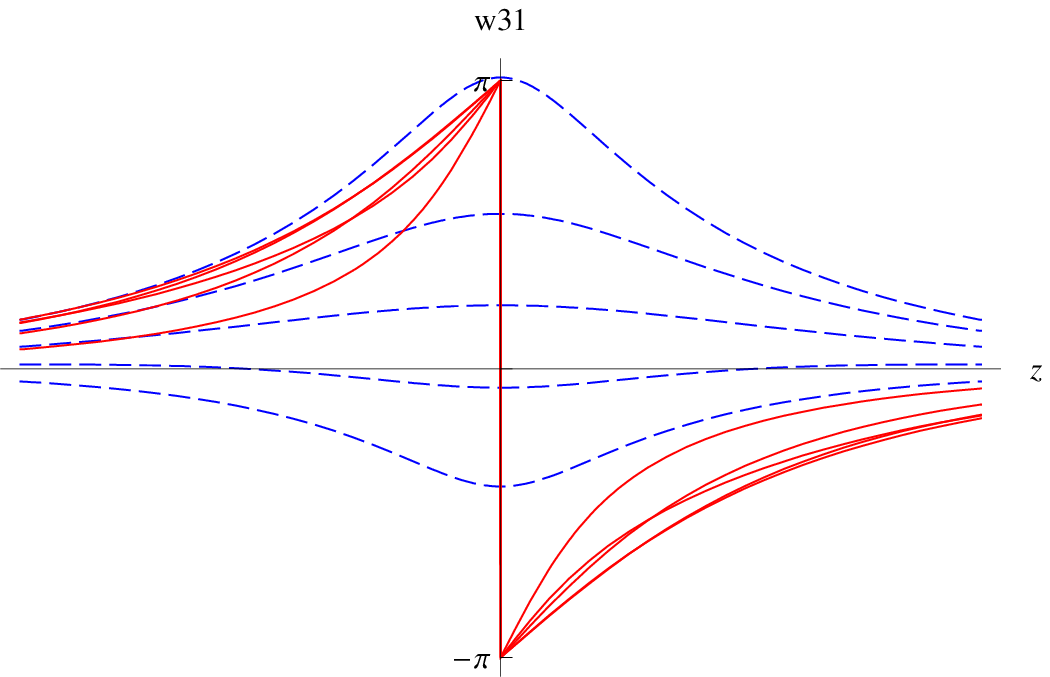}
\caption{A bunch of profile functions $w_{12}$ (left), $w_{23}$ (middle) and $w_{31}$ (right)
inside the $SU(3)$ string for five values of the parameter $\gamma$:
${\rm arg}(\gamma)=(4,5,6,7,8)\times (\pi/9)$.
The red solid curves display imaginary parts and the blue dashed curves display real
parts of $w_{12},w_{23},w_{31}$, respectively, as functions of the distance $z$ from the Wilson
loop plane. The string tension (the action) is identical for all five curves.}
\end{figure}

The action density also varies as function of ${\rm arg}(\gamma)$ but is periodic with a period
of $\frac{2\pi}{N}$. At $N$ points in the middle of the intervals \ur{argg1int}, namely at
${\rm arg}(\gamma)=\pi-\frac{(2m_1-1)\pi}{N}$, the action density is real; otherwise it is
generally complex. It is remarkable that the action itself, or the string tension, is real and
does not depend on $\gamma$. It means that ${\rm arg}(\gamma)$ is a new string Goldstone mode,
if one allows $\gamma$ to be a function of $2d$ string coordinates -- in addition to the usual
Goldstone modes associated with long-wave deformations of the string surface.

\subsection{Strings for higher representations, $k\geq 2$}

Wilson loop in the antisymmetric rank-$k$ tensor representation is a source for $k$ functions
$w_{m_1},\ldots ,w_{m_k}$ where the numbers $m_1<\ldots <m_k$ can lie anywhere on the circle
$(1,2,\ldots ,N)$. However, Toda equations have solutions not for all configurations of
$m_1\ldots m_k$. Configurations with no classical solutions presumably give much smaller
contributions to the Wilson loop at large areas than configurations that do generate solitons
as they are stationary points.

The strategy for finding pinned solitons corresponding to Wilson loops in higher representations
is simple: one takes the general solution \ur{pinned_soliton} at certain value of $k$ and varies
the phase of $\gamma$ from $\pi$ to $-\pi$. For any $k$ there will be {\em continuous} intervals of
${\rm arg}(\gamma)$ for which the functions $w^{(k)}_{m,m+1}(z)$ satisfy Toda equations \ur{UD4}
with a $\delta'(z)$ source in the r.h.s. corresponding to certain sets of numbers $m_1<m_2<\ldots < m_k$.
For all intervals of ${\rm arg}(\gamma)$, the pinned soliton action and hence the $k$-string
tension is given by \eq{sigmaM-k} and is thus degenerate in ${\rm arg}(\gamma)$.

We did not attempt to enumerate systematically the rapidly growing variety of solitons at arbitrary
$N$ and $k$. We find it more instructive to describe all pinned solitons of the $SU(6)$ group which is
sufficiently ``rich'' as it possesses non-trivial strings with $k=1,2$ and $3$.\\

For $k\!=\!1$ the solutions have been in fact given in subsection B: six equal-length intervals of
${\rm arg}(\gamma)\in\left(\pi,\frac{2\pi}{3}\right), \left(\frac{2\pi}{3},\frac{\pi}{3}\right),
\left(\frac{\pi}{3},0\right), \left(0,-\frac{\pi}{3}\right), \left(-\frac{\pi}{3},-\frac{2\pi}{3}\right),
\left(-\frac{2\pi}{3},-\pi\right)$ correspond to solutions with a single source placed at
$m_1=1,2,3,4,5,6$, respectively. In all cases the string tension is $\sigma(6,1)=2MT\,\sin\frac{\pi}{6}$.
\\

For $k\!=\!2$ three equal-length intervals ${\rm arg}(\gamma)\in\left(\pi,\frac{\pi}{3}\right),
\left(\frac{\pi}{3},-\frac{\pi}{3}\right),\left(-\frac{\pi}{3},-\pi\right)$ correspond to
the double $k\!=\!2$ sources at $m_1\!=\!1,m_2\!=\!4$; $m_1\!=\!2,m_2\!=\!5$; $m_1\!=\!3,m_2\!=\!6$,
respectively. In all cases the string tension is $\sigma(6,2)=2MT\,\sin\frac{2\pi}{6}$.
\\

For $k\!=\!3$ two equal-length intervals ${\rm arg}(\gamma)\in\left(\pi,0\right),
\left(0,-\pi\right)$ correspond to the triple $k\!=\!3$ sources at $m_1\!=\!1,m_2\!=\!3,m_3\!=\!5$
and $m_1\!=\!2,m_2\!=\!4,m_3\!=\!6$, respectively. In all cases the string tension is
$\sigma(6,3)=2MT\,\sin\frac{3\pi}{6}$.
\\

As a matter of fact $k=3,N=6$ is a particular case of the general rank-$k$ representation of
the $SU(2k)$ group. [Another example is the $k\!=\!1$ representation of the $SU(2)$ group,
which simultaneously is a particular case of a fundamental representation considered in
the previous subsection.] For all $k$ and $N=2k$ there are pinned solitons generated by
$k$ sources placed at $m_1\!=\!1,m_2\!=\!3,m_3\!=\!5,\ldots$ if ${\rm arg}(\gamma)\in(\pi,0)$,
and placed at $m_1\!=\!2,m_2\!=\!4,m_3\!=\!6,\ldots$ if ${\rm arg}(\gamma)\in(0,-\pi)$. The string tension
is given by \eq{sigmaM-k} where one puts $N=2k$, and is degenerate in ${\rm arg}(\gamma)$.
\\

To summarize this section, we have shown that to find the spatial Wilson loop averaged over
the ensemble of dyons, one needs to solve a chain of Toda equations with a $\delta'(z)$
source in the r.h.s. We have solved those equations for any $N$ and Wilson loop representation $k$,
finding pinned solitons in the transverse direction to the loop surface. The solutions
generalize the famous double-layer solutions for the string in the $3d$ Georgi--Glashow model
by Polyakov~\cite{Polyakov77}. The resulting `magnetic' string tension is proportional to
$\sin\frac{\pi k}{N}$ and coincides exactly with the `electric' string tension found in
Section V from the correlators of the Polyakov lines. We have observed that the Toda equations
with a source allow a continuous set of solutions for the string profile, characterized
by a phase ${\rm arg}(\gamma)\in(-\pi,\pi)$, all with the same string tension. It means that
in addition to the usual Goldstone modes related to deformations of the string surface,
there must be an extra Goldstone mode related to the string profile. Therefore, the
string theory is more complicated than given by the standard Nambu--Goto action, which may
have important implications both for theory and phenomenology.

\newpage

\section{Summary}

Generalizing previous work on the subject, we have written down the metric of
the moduli space for an arbitrary number of $N$ kinds of dyons in the pure
$SU(N)$ gauge theory. Assuming that it is mainly the metric and not the
fluctuation determinant about dyons that defines the ensemble of interacting
dyons, we have presented the grand partition function of the ensemble
(where the number of particles is not fixed beforehand but found from the
minimum of the free energy at given temperature) as a path integral over
$N\!-\!1$ Abelian electric potentials $v_m$ and their duals $w_m$, as well as
over $N\!-\!1$ ghost fields $\chi^\dagger_m,\chi_m$. The resulting quantum field
theory of those fields turns out to be exactly solvable owing to the
cancelation between boson and ghost loops -- a feature similar to that observed
in supersymmetric theories. It enables one to make exact statements about the
dyon ensemble: to find its free energy and correlation functions.

The free energy appears to have the minimum at the ``maximal non-trivial''
holonomy corresponding to the confining zero value of the average Polyakov
line. Calculating the correlation functions of Polyakov lines in various
$N$-ality $=k$ representations (where $k=1,\ldots\,N\!-\!1$) we find the
asymptotic linear confining potential with the $k$-string tension proportional
to $\sin\frac{\pi k}{N}$, the coefficient being calculated through the Yang--Mills
scale parameter $\Lambda$ and the 't Hooft coupling $\lambda$. The actual value
of $\lambda$ has to be determined self-consistently at the 2-loop level
not considered here. Taking $\lambda=\frac{1}{4}$ compatible with phenomenology
we observe a reasonable agreement of the estimated deconfinement temperature $T_c$,
the string tension $\sigma$, the gluon condensate and the topological susceptibility
with what is known from lattice simulations and phenomenology. A more robust
ratio $T_c/\sqrt{\sigma}$ independent of $\Lambda$ and $\lambda$ is in surprisingly
good agreement with the lattice data taken at $N=3,4,6$ and $8$, given the
approximate nature of the model.

We have also calculated the string tension from the area law for the average of
{\em spatial} Wilson loops for any $N$ and $k$ in our dyon ensemble.
The spatial (`magnetic') string tension coincides with the
`electric' string tension found from Polyakov lines for all $N$ and $k$.
We find this coincidence interesting as it indicates the
restoration of Lorentz symmetry at low temperatures. Since the formalism used
is 3-dimensional at finite temperatures, the restoration of Lorentz symmetry at
$T\!\to\!0$ is by no means trivial.

We do not pretend we have answered all the questions and obtained a realistic
confining theory as we have ignored essential ingredients of the
full Yang--Mills theory, enumerated in the Introduction. Our aim was to
demonstrate that the integration measure over dyons has a drastic, probably a
decisive effect on the ensemble of dyons, that the ensemble can be mathematically
described by an exactly solvable field theory in three dimensions, and that the
resulting semiclassical vacuum built of dyons has many features expected for
the confining pure Yang--Mills theory.

\vskip 0.3true cm

\noi
{\large\bf Acknowledgements}\\

We thank Nikolaj Gromov and Alexei Yung for very helpful discussions.
D.D. acknowledges useful discussions with Ulf Lindstr\"om, Maxim Zabzine and Konstantin
Zarembo and their kind hospitality during the visit to the Theoretical Physics Department
at the University of Uppsala.
This work has been supported in part by the Russian Government grants RFBR-06-02-16786
and RSGSS-5788.2006.2.


\begin{thebibliography}{99}

\bibitem{BPS}
E.B.~Bogomolny, Yad. Fiz. 24, 861 (1976) [Sov. J.
Nucl. Phys. {\bf 24}, 449 (1976)];\\
M.K.~Prasad and C.M.~Sommerfield, Phys. Rev. Lett. {\bf 35}, 760 (1975).

\bibitem{LeeYi}
K.~Lee and P.~Yi, Phys. Rev. {\bf D56}, 3711 (1997), ArXive: hep-th/9702107.

\bibitem{DHKM-99}
N.M.~Davies, T.J.~Hollowood, V.V.~Khoze and M.P.~Mattis,
Nucl. Phys. {\bf B559}, 123 (1999), ArXive: hep-th/9905015.

\bibitem{KvB}
T.C.~Kraan and P.~van Baal, Phys. Lett. {\bf B428}, 268 (1998), ArXive: hep-th/9802049;
Nucl. Phys. {\bf B533}, 627 (1998), ArXive: hep-th/9805168.

\bibitem{LL}
K.~Lee and C.~Lu, Phys. Rev. {\bf D58}, 025011 (1998), ArXive: hep-th/9802108.

\bibitem{DGPS}
D.~Diakonov, N.~Gromov, V.~Petrov and S.~Slizovskiy, Phys. Rev. {\bf D70}, 036003 (2004),
ArXive: hep-th/0404042; \\
D.~Diakonov, in Continuous Advances in QCD 2004, ed. T.~Ghergetta,
World Scientific (2004) p. 369, ArXive: hep-ph/0407353.

\bibitem{BPST}
A. Belavin, A. Polyakov, A. Schwartz and Yu. Tyupkin, Phys. Lett. {\bf 59}, 85 (1975).

\bibitem{HS}
B.J. Harrington and H.K. Shepard, Phys. Rev. {\bf D17}, 2122 (1978);
Phys. Rev. {\bf D18}, 2990 (1978).

\bibitem{GPY}
D.J.~Gross, R.D.~Pisarski and L.G.~Yaffe, Rev. Mod. Phys. {\bf 53}, 43 (1981).

\bibitem{NW}
N.~Weiss, Phys. Rev. {\bf D24}, 475 (1981);  Phys. Rev. {\bf D25}, 2667 (1982);\\
D.~Diakonov and M.~Oswald, Phys. Rev. {\bf D70}, 105016 (2004), ArXive: hep-ph/0403108.

\bibitem{D02}
D. Diakonov, Prog. Part. Nucl. Phys. {\bf 51}, 173 (2003), ArXive: hep-ph/0212026.

\bibitem{GS-SUN}
N.~Gromov and S.~Slizovkiy, in preparation.

\bibitem{ILM}
C.~Callan, R.~Dashen and D.~Gross, Phys. Rev. {\bf D17}, 2717 (1978);\\
E.M.~Ilgenfritz and  M.~M\"uller-Preussker, Nucl. Phys. {\bf B184}, 443 (1981);\\
E.~Shuryak, Nucl. Phys. {\bf B203}, 93 (1982);\\
D.~Diakonov and V.~Petrov, Nucl. Phys. {\bf B245}, 259 (1984).

\bibitem{Nogradi}
F.~Bruckmann, D.~N\'ogr\'adi and P. van Baal, Nucl. Phys. {\bf B698}, 233
(2004), ArXive: hep-th/0404210;\\
D.~N\'ogr\'adi, PhD thesis, Leiden University (2005), ArXive: hep-th/0511125.

\bibitem{AH}
M.F.~Atiyah and N.J.~Hitchin, Phys. Lett. {\bf A107}, 21 (1985).

\bibitem{GM1}
G.W.~Gibbons and N.S.~Manton, Nucl. Phys. {\bf B274}, 183 (1986).

\bibitem{GM2}
G.W.~Gibbons and N.S.~Manton, Phys. Lett. {\bf B356}, 32 (1995).

\bibitem{Ilgenfritz}
P.~Gerhold, E.M.~Ilgenfritz, M.~M\"uller-Preussker, Nucl. Phys. {\bf B760}, 1
(2007); {\it ibid.} {\bf B774}, 268 (2007).

\bibitem{LWY}
K.M.~Lee, E.J.~Weinberg and P.~Yi,
Phys. Rev. {\bf D54}, 1633 (1996), ArXive: hep-th/9602167;
see also Ref.~\cite{LeeYi}.

\bibitem{Kraan}
T.C.~Kraan, Commun. Math. Phys. {\bf 212}, 503 (2000), ArXive: hep-th/9811179;
PhD Thesis, Leiden University (2000)
[available at http://www.lorentz.leidenuniv.nl/vanbaal/HOME/PUBL/kraan.ps].

\bibitem{ADHM}
M.F.~Atiyah, V.G.~Drinfeld, N.J.~Hitchin and Yu.I.~Manin, {\bf 65A}, 185 (1978);\\
N.H.~Christ, E.J.~Weinberg and N.K.~Stanton, Phys.~Rev. {\bf D18}, 2013 (1978);\\
E.~Corrigan, P.~Goddard and S.~Templeton, Nucl.~Phys. {\bf B151}, 93 (1979).

\bibitem{Nahm80}
W.~Nahm, Phys.~Lett. {\bf B90} (1980) 413.

\bibitem{KvBSUN}
T.C.~Kraan and P.~van Baal,
Phys. Lett. {\bf B435}, 389 (1998), ArXive: hep-th/9806034.

\bibitem{DG05}
D.~Diakonov and N.~Gromov, Phys. Rev. {\bf D72}, 025003 (2005), ArXive:
hep-th/0502132.

\bibitem{Guadagnini}
E.~Guadagnini, Nucl. Phys. {\bf B236}, 35 (1984).

\bibitem{DPmon}
D. Diakonov and V. Petrov, in: Non-perturbative approaches to Quantum
Chromodynamics, Proc. int. workshop at ECT*, Trento, 1995, ed. D. Diakonov,
Gatchina (1995) p. 36, ArXive: hep-th/9606104.

\bibitem{Bernard}
G. 't Hooft, Phys. Rev. {\bf D14}, 3432 (1976); Erratum: {\it ibid.} {\bf D18}, 2199
(1978);\\
C.~Bernard, Phys. Rev. {\bf D19}, 3013 (1979).

\bibitem{GP}
G.W.~Gibbons and C.N.~Pope, Commun. Math. Phys. {\bf 66}, 267 (1979).

\bibitem{AHbook}
M.F.~Atiyah and N.J.~Hitchin, {\it The Geometry and Dynamics of Magnetic
Monopoles}, Princeton University Press (1988).

\bibitem{footnote1}
The definition of the metric tensor is the matrix composed of zero modes' overlaps.
At large separations between dyons, the four approximately zero modes per dyon
are $F^{(i)}_{\mu\alpha}, \;\alpha=1,2,3,4,$ where $F^{(i)}$ is the field
strength of the $i^{\rm th}$ dyon whose scale parameter $\nu$ has to be
adjusted by the Coulomb field of other dyons. This is why the first
correction to the metric is always Coulomb-like. It is remarkable that for
different-kind dyons there are no further corrections.

\bibitem{Gromov}
N.~Gromov, ArXive: hep-th/0701192.

\bibitem{DP1}
D.~Diakonov and V.~Petrov, Ref.~\cite{ILM}.

\bibitem{Berezin}
F.A.~Berezin, {\it Second Quantization Method}, Nauka, Moscow (1965) (in Russian).

\bibitem{Polyakov77}
A.~Polyakov, Nucl. Phys. {\bf B120}, 429 (1977).

\bibitem{footnote15}
In principle, one can think of a quantum anomaly leading to an uncomplete
cancellation of boson and ghost determinants, as it happens in some
supersymmetric examples~\cite{Shifman-Yung}. We do not see a reason for
the anomaly in our case, however. Therefore, unless proven otherwise, we shall
assume that the cancellation is exact.

\bibitem{Shifman-Yung}
M.~Shifman and A.~Yung, ArXive: hep-th/0703267.

\bibitem{Zhitnitsky}
S.~Jaimungal and A.R.~Zhitnitsky, ArXive: hep-ph/9905540;
A.R.~Zhitnitsky, ArXive: hep-ph/0601057.

\bibitem{Toda}
M.~Toda, Phys. Rep. {\bf 18C}, 1 (1975). It should be noticed, however, that
the classic Toda lattice is of the ``hyperbolic'' type, {\it i.e.} the sign
in the definition of ${\cal F}$ \ur{calF} is opposite. Here the Toda lattice
is of the ``elliptic'' type.

\bibitem{footnote2}
Temperature dependence of the string tension appears when one adds {\it e.g.}
the perturbative potential energy into consideration. Since its relative contribution
is $\sim T^4$ one expects that it modifies seriously \eq{sigmaE} only close to the
phase transition temperature where the electric string tension vanishes.

\bibitem{Teper}
B.~Lucini, M.~Teper and U.~Wenger, JHEP {\bf 0401}, 061 (2003), ArXive: hep-lat/0307017;\\
B.~Lucini, M.~Teper and U.~Wenger,  ArXive: hep-lat/0502003.

\bibitem{TeperQT}
B.~Lucini and M.~Teper, ArXive: hep-lat/0103027.

\bibitem{DP-SUSY}
D.~Diakonov and V.~Petrov, Phys. Rev. {\bf D67}, 105007 (2003), ArXive: hep-th/0212018.

\bibitem{Shifman}
M.~Shifman, Acta Phys. Polon. {\bf B6}, 3805 (2005), ArXive: hep-th/0510098.

\bibitem{sine}
M.R.~Douglas and S.H.~Shenker, Nucl. Phys. {\bf B447}, 271 (1995);\\
A.~Hanany, M.~Strassler and A.~Zaffaroni, Nucl. Phys. {\bf B513}, 87 (1998),
ArXive: hep-th/9707244.

\bibitem{footnote3}
In our linearized approach, there are combinatorial contributions
$\sim\exp\left(-|{\bf z}_1\!-\!{\bf z}_2|M(\sqrt{{\cal M}^{(l_1)}}
+\sqrt{{\cal M}^{(l_2)}}+\ldots)\right)$ to the correlator, stemming from
expanding the exponent in \eq{corr3} and its analogs for arbitrary $k$ in
higher powers of
$\exp\left(-|{\bf z}_1\!-\!{\bf z}_2|M\sqrt{{\cal M}^{(l)}}\right)$. Starting
from $k\!\geq\!3$ there is a danger that a combination of lower eigenvalues
may give a larger contribution to the correlator than a single contribution
of ${\cal M}^{(k)}$, for example, $2\surd {\cal M}^{(1)}<\surd {\cal M}^{(3)}$.
For this particular case we have proved that the square of the ${\cal M}^{(1)}$
contribution decouples from the $k\!=\!3$ correlator for all $N$. However,
we have not proved it for all $k$ and all possible combinations
of eigenvalues whose sum is less than $\surd{\cal M}^{(k)}$: probably certain
generalizations of the orthogonality relations \ur{ortho} are responsible for
it. A more appropriate approach would be to solve the full non-linear equations.
We note that combinatorial contributions leading to {\em larger} string tensions
are generally present and can be considered as string excitations.

\bibitem{DelDebbio-k}
L.~Del Debbio, H.~Panagopoulos, P.~Rossi and E.~Vicari, JHEP {\bf 0201}, 009 (2002),
ArXive: hep-th/0111090.

\bibitem{Teper-k}
B.~Lucini, M.~Teper and U.~Wenger, JHEP {\bf 0406}, 012 (2004), ArXive: hep-lat/0404008.


\bibitem{Toda-sol}
T.J.~Hollowood, ArXive: hep-th/9110010.


\end{thebibliography}
\end{document}